\documentclass[sigconf, nonacm]{acmart}

\usepackage[english]{babel}
\usepackage{verbatim}
\usepackage{diagbox}
\usepackage{color}
\usepackage{multirow}
\usepackage{url}
\usepackage{graphicx}
\usepackage{enumitem} 
\usepackage{xspace}
\usepackage{amsmath}
\usepackage{tikz}
\usepackage{tabularx}
\usepackage{balance}
\usepackage{soul}
\usepackage{amsthm}
\usepackage{subcaption}
\usepackage{array}
\newcommand{\PreserveBackslash}[1]{\let\temp=\\#1\let\\=\temp}
\newcolumntype{C}[1]{>{\PreserveBackslash\centering}p{#1}}
\newcolumntype{R}[1]{>{\PreserveBackslash\raggedleft}p{#1}}
\newcolumntype{L}[1]{>{\PreserveBackslash\raggedright}p{#1}}

\usepackage{mathtools}
\DeclarePairedDelimiter\ceil{\lceil}{\rceil}

\newcommand{\sysname}{MiCS\xspace}

\newcommand{\parabf}[1]{\medskip\noindent\textbf{#1}}

\newcommand{\cut}[1]{}

\newcommand{\zhen}[1]{\textcolor{black}{#1}}

\begin{document}
\title{MiCS: Near-linear Scaling for Training Gigantic Model on Public Cloud}

\author{Zhen Zhang}
\authornote{Work mostly done as an intern at Amazon Web Services.}
\affiliation{%
  \institution{Johns Hopkins University}
}
\email{zzhen1@jhu.edu}

\author{Shuai Zheng}
\affiliation{
  \institution{Amazon Web Services}
}
\email{shzheng@amazon.com}

\author{Yida Wang}
\affiliation{
  \institution{Amazon Web Services}
}
\email{wangyida@amazon.com}

\author{Justin Chiu}
\affiliation{
  \institution{Amazon}
}
\email{justchiu@amazon.com}

\author{George Karypis}
\affiliation{
  \institution{Amazon Web Services}
}
\email{gkarypis@amazon.com}

\author{Trishul Chilimbi}
\affiliation{
  \institution{Amazon}
}
\email{trishulc@amazon.com}

\author{Mu Li}
\affiliation{
  \institution{Amazon Web Services}
}
\email{mli@amazon.com}

\author{Xin Jin}
\affiliation{
  \institution{Peking University}
}
\email{xinjinpku@pku.edu.cn}

\begin{abstract}
Existing general purpose frameworks for gigantic model training, i.e., dense models with billions of parameters, cannot scale efficiently on
cloud environment \zhen{with various networking conditions} due to large communication overheads.
In this paper, we propose \sysname, which \underline{Mi}nimizes the \underline{C}ommunication \underline{S}cale to bring down communication overhead.
Specifically, by decreasing the number of participants in a communication collective, 
\sysname can utilize heterogeneous network bandwidth,
reduce network traffic
over slower links, \zhen{
reduce the latency of communications for maintaining high network bandwidth utilization,
} and amortize expensive global gradient synchronization overhead.
Our evaluation on AWS shows that the system throughput of \sysname is
up to 2.89$\times$ that of the state-of-the-art large model training systems.
\sysname achieves near-linear scaling efficiency, which is up to
1.27$\times$ that of DeepSpeed. \sysname allows us to train a proprietary model with 100 billion parameters on 512 GPUs with 99.4\% weak-scaling
efficiency, and it is able to saturate over 54.5\% theoretical computation power
of each GPU on a public cloud with less GPU memory and more restricted networks than DGX-A100
clusters.

\end{abstract}


\maketitle




\section{Introduction}
\label{sec:introduction}

There is a growing body of research showing that large Deep Learning (DL) models deliver superior accuracy in areas such as natural language processing (NLP)~\cite{megatron_lm, bert}, speech recognition (SR)~\cite{zhang2020pushing, chung2021w2vbert, chan2021speechstew}, and computer vision (CV)~\cite{wideresnet,dai2021coatnet, zhai2021scaling}.
This has resulted in a more than 1000$\times$ increase in the size of the DL models that are commonly trained with many of them having several hundred billion parameters.
The high computational requirement associated with training DL models has led to effective and simple parallelization approaches based on data parallelism (DP)~\cite{pytorch, sergeev2018horovod,tensorflow, byteps, ps-lite}. However, many of these approaches cannot be applied for training gigantic DL models, as their memory requirements exceed the amount of GPU memory.

%
A common way to train gigantic DL models is to use model-parallelism (MP) that decomposes the computation across the devices by partitioning the neural network architecture (i.e., the model). As a result of this network partitioning, the \emph{model states} (i.e., the memory storing the model parameters, gradients, and optimizer states) are also partitioned across the devices, and as such it overcomes DP's memory-related limitations. Unfortunately, existing MP frameworks require users to substantially modify the logic of their training code and add specific primitives~\cite{shazeer2018meshtensorflow, flexflow,lepikhin2020gshard, megatron_lm, narayanan2019pipedream}. In addition, many of the MP frameworks are specifically designed for certain types of neural network architectures~\cite{megatron_lm, DLRM19} and cannot be directly used for arbitrary architectures.
However, the idea of partitioning the model states across different devices is essential for enabling large model training and was recently incorporated into DP by the development of ZeRO~\cite{rasley2020deepspeed}. ZeRO, which is implemented in distributed systems DeepSpeed~\cite{rasley2020deepspeed} and FairScale~\cite{fairscale}, evenly partitions the model states across the entire training cluster, enabling the training of very large models while retaining DP's simplicity, ease of use, and generality.

ZeRO was designed for clusters using nodes based on NVIDIA's DGX-2 or DGX-A100 multi-GPU systems~\cite{megatron_lm_3d,ms_zero}. These nodes are connected via high-bandwidth low-latency InfiniBand leading to clusters whose intra- and inter-node GPU-to-GPU bandwidth is nearly balanced (intra-node bandwidth is about 3$\times$ faster than inter-node). 
ZeRO takes advantage of this balanced network to treat all GPU devices of the cluster equivalently and to partition the model states across the entire cluster. As a result, whenever 
during the forward or backward phase a parameter tensor is required for the computations, a collective communication operation needs to be performed that involves all devices of the entire cluster (\S\ref{sec:large_model_training}). 
\zhen{Training clusters in public cloud environment are not always equipped with high-speed InfiniBand networks as DGX nodes have. For example, cloud instances with V100 GPUs are typically paired with 100Gbps networks~\cite{azure-gpu-ncv3,azure-gpu-ndv2,gcloud-gpu-bandwidths,awsp3}, in which case the bandwidth is less balanced (intra-node bandwidth is about 24$\times$ faster than inter-node). In such scenarios, ZeRO is not well suited. By treating these devices equivalently and not accounting for the heterogeneous and hierarchical nature of the inter-node network, ZeRO fails to take advantage of the faster intra-node networks. Moreover, by partitioning the model states across the entire cluster, even when the model states can fit in the memory of a subcluster, ZeRO unnecessarily incurs high communication cost of collective communications, because of the low effective bandwidth caused by high algorithmic latency. And the communication overhead of ZeRO grows larger as the size of the cluster scales up (\S\ref{sec:comm_overheads}).}

To surmount the aforementioned challenges, we propose \sysname, following a core design principle: to reduce the number of communicating participants, i.e., \emph{communication scale}, as much as possible.
By minimizing the scale, \sysname reduces the latency and the data volume transmitted over slow inter-node links.
We design and implement three components to realize the design principle for reducing communication overheads.
\begin{itemize}[leftmargin=*]

  \item \emph{Scale-aware model partitioning.} 
  Instead of using all devices as a single group for holding the model states, \sysname divides all devices into partition groups. Each group holds a complete copy of the model states. Within each group, the model states are partitioned. Thus most frequent parameter gatherings are operated at the \emph{scale} of each group (\S\ref{sec:sys_design_topo_aware}).

  \item \emph{Hierarchical communication strategy.} Hierarchical communication allows us to parallelize multiple inter-node collective communications and reduce the scale of each collective communication. 
  It reduces the aggregated traffic over the inter-node links, thus leading to lower communication cost (\S\ref{sec:sys_design_hierarchy_sync}).

  \item \emph{2-hop gradient synchronization.} Unlike ZeRO that synchronizes gradients over all devices for each micro-step, \sysname only synchronizes gradients within the partition group until the gradient accumulation boundary is reached. At the gradient accumulation boundary, gradients are synchronized across the partition groups. 
  As a result, \sysname reduces the synchronization cost significantly by amortizing the cost to multiple micro-steps. (\S\ref{sec:sys_design_grad_sync}).

\end{itemize}

Our thorough evaluation shows significant system throughput and scaling efficiency improvement of \sysname on public clouds like AWS. On V100 GPU clusters with 100Gbps network, the system throughput of \sysname is 2.89$\times$ larger than that of DeepSpeed, which is the state-of-the-art DP framework for large model training. On A100 GPU clusters with 40GB memory per GPU and 400Gbps networks, \sysname is up to 2.74$\times$ as fast as DeepSpeed. Compared to Megatron-LM-3D, a state-of-the-art system specialized for training Transformer models, \sysname achieves up to 30.1\% better throughput.
\sysname gets near-linear (e.g., 99.4\%) weak scaling efficiency in the cloud, which is up to 27\% better than DeepSpeed.
\sysname has been deployed to train a proprietary model with 100 billion (B) parameters, saturating over 170 TeraFLOP/s (TFLOPS) per A100 GPU \zhen{with activation checkpointing} at the scale of 512 GPUs.

In summary, this paper makes the following contributions. 
\begin{itemize}[leftmargin=*]
    \item We identify the root problem---overwhelming communication overhead---that prevents DP-based model partitioning from efficiently scaling out on clusters with 100Gbps or 400Gbps network interfaces with relatively higher latency than InfiniBand~\cite{ziegler2022efa-dbms-latency}.
    \item We design and implement a system \sysname that minimizes the communication scale to reduce the communication overhead. 
    \item We evaluate \sysname thoroughly to justify the benefits of minimizing communication scale on clusters with up to 512 GPUs.
\end{itemize}

\section{Background and Motivation}
\label{sec:background_movtivation}
In this section, we briefly review deep learning model training (\S\ref{sec:model_training}) and how existing works tackle the large model training challenges (\S\ref{sec:large_model_training}), and discuss its major limitation in the context of public clouds (\S\ref{sec:comm_overheads}). We then present the intuition that motivates our design (\S\ref{sec:motivation}).

\subsection{Model Training}
\label{sec:model_training}
Deep learning model training process mainly consists of three phases, i.e., forward computation, backward computation, and parameter updating.
In order to train the model faster, we can harness the computing power of multiple machines. A gradient synchronization step is performed before updating the model parameters to ensure all workers will use the same set of parameters to evaluate the incoming new training samples.

Deep learning model training is memory consuming as it needs to hold the model states including model parameters, gradients from backward computation, and optimizer states for parameter updating.
Because of the limited on-device memory resource, \emph{activation checkpointing} and \emph{gradient accumulation} are typically enabled. Activation checkpointing discards the activation outputs from the forward phase and requires activation recomputation in the backward phase. Gradient accumulation divides one large data batch into multiple small micro-batches to reduce the memory footprint of storing activation outputs.
However, for models with billions of parameters, these two techniques alone are not sufficient. Many solutions targeting at gigantic model training are thus proposed. 

\begin{figure*}
    \centering
    \includegraphics[width=\linewidth]{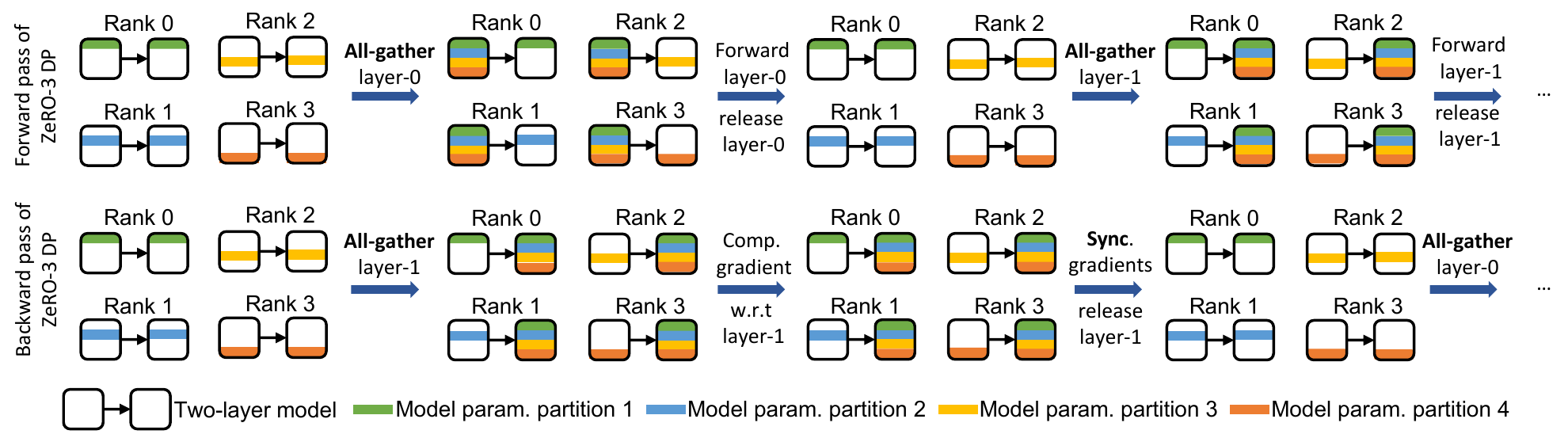}
    \caption{\zhen{Forward and backward passes of ZeRO-3 Data Parallelism on a cluster with four devices; for brevity, only the model parameter states are shown in the figure; ``Comp.'': compute, ``Sync.'': synchronize, ``param.'': parameter; communications are marked with bold font; a two-layer model is used here for illustration purposes.}}
    \label{fig:partitioned-dp-zero3}
\end{figure*}

\subsection{Gigantic Model Training}
\label{sec:large_model_training}

\zhen{In this paper, we use the term ``gigantic model'' to refer to those Deep Neural Network models that consist of billions of densely connected parameters, which means both the size of the model and the per-sample computation of the model, i.e., floating-point operations (FLOPs), are ``gigantic''. Nowadays, the commonly adopted models that fall into this category are transformer-based models~\cite{megatron_lm_3d,ms_zero,gpt2, gpt3, opt175, ds-megatron-530b} and latest wide computer vision models~\cite{wideresnet}. }

Traditionally, developers use model-parallel (MP) distributed training for gigantic model training. The basic idea is to distribute the model parameters and computations across multiple devices for each training sample. Thus, the memory for storing model states is also distributed across devices. This way of distributing computations comes with issues. Tensor model parallelism as one MP method requires lots of communications during computation~\cite{megatron_lm}. On the other hand, pipeline MP strategy is advocated with smaller communication overheads, but it suffers from pipeline bubbles and causes under-utilization. Besides, MP solutions are not directly compatible with common frameworks like PyTorch or Tensorflow, and they require non-trivial engineering effort from the user side. Lastly, some of the MP designs~\cite{megatron_lm_3d, DLRM19} are model-specific solutions, making them hard to generalize.

Compared to the MP solutions, ZeRO~\cite{ms_zero} powered data-parallel (DP) solutions are general to various models and do not require model refactoring. ZeRO partitions the \emph{model states} onto all devices on the cluster to reduce the memory consumption on each device. ZeRO has three different stages, corresponding to three different levels of memory reduction:
ZeRO-1 partitions optimizer states only;
ZeRO-2 partitions gradients and optimizer states;
ZeRO-3 partitions all three states\zhen{, i.e., parameters, gradients and optimizer states,} evenly across all devices on the training cluster.
The full-fledged ZeRO allows us to train the extremely large models when we have a large enough cluster. However, we have to pay communication costs for gathering model parameters during both forward and backward.
\zhen{Figure~\ref{fig:partitioned-dp-zero3} illustrates the forward and backward passes in ZeRO-3 powered DP, in which the parameters of each layer are partitioned across all the ranks. Here we use the same convention in the high-performance computing (HPC) community where we use a rank number to identify a computing device. Before computing the activations or gradient for a layer, all parameters of this layer are gathered back by \texttt{all-gather} communication. After computing the gradients on each rank with its own part of the data, the gradients are synchronized and partitioned across all ranks using \texttt{reduce-scatter} communication, which aggregates gradients among all ranks and partitions the gradients at the same time. The gradient partition is necessary for gigantic models with billions of parameters due to the limited memory on each rank. }

ZeRO-Offload~\cite{ren2021zerooffload} and ZeRO-Infinity~\cite{rajbhandari2021zeroinfinity} are two extensions to ZeRO-3. These two systems offload model parameters, gradients, and optimizer states to CPU memory and NVMe SSDs. Both systems suffer from the same communication overheads as ZeRO-3, which will be discussed in the next subsection.

\subsection{Communication Overhead}
\label{sec:motivation_communication_bottleneck}
\label{sec:comm_overheads}

\begin{figure}[t]
    \centering
    \includegraphics[width=0.85\linewidth]{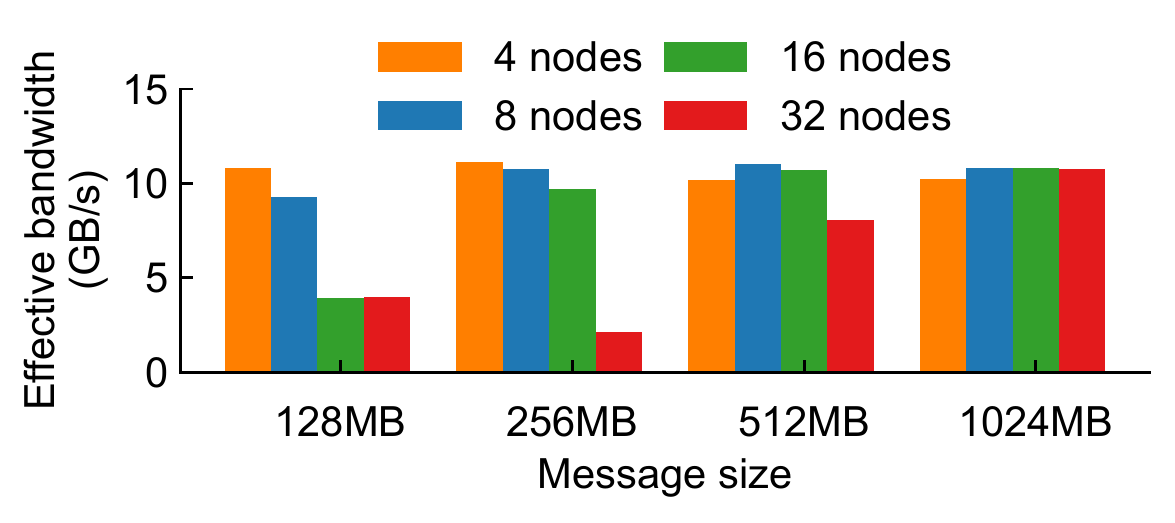}
    \caption{Effective bandwidths measured with all-gather.}
    \label{fig:motivation_bandwdith_utilization}
\end{figure}
ZeRO's model state partitioning mechanism results in the heavy use of collective communication for gathering model states \zhen{, which is demonstrated in Figure~\ref{fig:partitioned-dp-zero3}. }
\zhen{Specifically, ZeRO-3 transmits $3(n-1)M/n$ bytes~\cite{ms_zero} in forward and backward passes, where $M$ denotes the size of the parameters of the model in bytes, $n$ denotes the number of devices. The transmitted data volume is as large as tens to hundreds of gigabytes for models with tens to hundreds of billions of parameters. The cost of transmitting these data crossing the entire cluster cannot be easily hidden via pipelining the communication and computation. }
Our timeline measurement shows that for a BERT model with 10B parameters, parameter gathering could take 2.85$\times$ more time than computation in forward pass.
Similar expensive communications also exist in the backward computation and gradient synchronization, which hurts the performance of ZeRO especially when the network bandwidth between devices is less preferable.

\zhen{There are two main factors that contribute to the costly communications when using ZeRO-3 on the cloud. Firstly, at the hardware level, many of the available internode network bandwidths and latency of cloud-based GPU clusters are not as good as DGX systems~\cite{gcloud-gpu-bandwidths,awsp3,azure-gpu-ndv2, ziegler2022efa-dbms-latency}. Moreover, unlike on-premise clusters, the network topology of cloud-based clusters is out of users' control, which could negatively impact the network performance~\cite{luo2020plink, luo2021cloud}. 
}
Secondly, at the algorithmic side, the latency of collective algorithms for communication has a positive correlation with the communication scale and the startup time for transmission~\cite{CollectiveCommunicationTheory_chan2007}\footnote{The latency of tree algorithms is bounded with $\ceil*{\log_2(p)}\alpha$, and the ring algorithms have a latency term $2p\alpha$, where $p$ denotes the number of participants and $\alpha$ denotes the startup time for transmission, \S7.1.7 in \cite{CollectiveCommunicationTheory_chan2007}}.
Therefore, as the scale grows, the latency becomes more significant and hurts the performance of communication at a large scale.
In addition, previous studies~\cite{zhang2020network} suggest that network bandwidth may not be the performance bottleneck of distributed model training.
Based on our measurements, as the cluster size grows, we need larger message sizes to saturate the bandwidth. Figure~\ref{fig:motivation_bandwdith_utilization} shows that small-size message such as 128MB obtains poor bandwidth utilization on 16 and 32 nodes.

In practice, we may not be able to always communicate large messages due to the memory constraint. Instead, it is better to control the communication scale to improve the bandwidth utilization, especially on the cloud with less favorable network conditions.

\subsection{Motivation}
\label{sec:motivation}
As mentioned in \S\ref{sec:motivation_communication_bottleneck}, the frequent communication among all devices significantly hampers the training performance of ZeRO powered DP solutions. This motivates us to design a new system that reduces the cost of communications while preserving the generality and usability advantages.
We found the communication overhead can be effectively reduced by shrinking the communication scale, i.e., reducing the number of participants in a collective communication. With the reduced communication scale, the majority of communications are restricted to a smaller group of devices. This allows us to maintain high bandwidth utilization in communications for various sized messages, as shown in Figure~\ref{fig:motivation_bandwdith_utilization}.
In addition, because the transmitted data volume is positively correlated to the number of participants, reducing the communication scale reduces the data volume. We give detailed descriptions of our methodology in the next section.

\section{\sysname Design}
\label{sec:sys_design}

\sysname is designed for training large models on the public cloud.
The overarching goal of \sysname's design is to reduce the scale of communication. The reduced scale allows us to exploit heterogeneous network bandwidth, and to reduce the network traffic transmitted over slow links. To effectively reduce the communication scale, we propose
three components named \emph{small-scale model partitioning}, \emph{hierarchical communication} and \emph{2-hop gradient synchronization}. For each of them, we explain the motivation, the methodology, and the analysis of our design.

\subsection{Notation}
We define the notations used in this section as follows:
\begin{itemize}[leftmargin=*]
    \item $n$: Number of devices or ranks in the cluster.
    \item $k$: Number of devices on each computational node.
    \item $M$: Size of a model in bytes.
    \item $p$: Number of devices for holding a model replica.
    \item $s$: Number of micro-steps.
    \item $B_{g}$: Effective communication bandwidth among devices belonging to the group $g$. We define the effective communication bandwidth as the bandwidth measured using collective communication. Effective communication bandwidth takes algorithm latency into account. Thus it is smaller than the theoretical bandwidth of hardware specification. For a fixed message size, when the number of nodes increases, the effective bandwidth shrinks.
    \item $C$: Time cost.
\end{itemize}


\subsection{Scale-aware Model Partitioning}
\label{sec:sys_design_topo_aware}
\begin{figure}[t]
  \centering
  \includegraphics[width=0.95\linewidth]{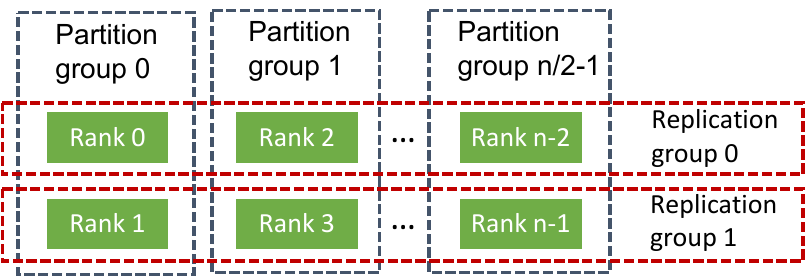}
  \caption{
    Model training on $n$ devices,
  with every 2 devices together holding a copy of the model states. Each \emph{partition group} maintains a copy of the entire model states, and devices in a \emph{replication group} hold the same part of model states.}
  \label{fig:sys_design_partition}
\end{figure}

Partitioning model states across all the devices causes significant communications overheads during training. Such communication overheads scale with the number of participants in a single collective communication (\S\ref{sec:motivation_communication_bottleneck}).
To reduce the communication overheads, we consider distributing the model states over a subset of devices to reduce the scale of the communication. A modern single computing device typically has tens of gigabytes of memory, and tens of them provide sufficient memory for a model with tens of billions of parameters. For example, a model with 10 billion parameters takes about 160GB of memory when training with Adam optimizer using mixed-precision. Partitioning the model states across 8 V100 (32GB) GPUs is already more than enough. By using 8 V100 GPUs instead of all the devices for holding one model states replica, we can effectively reduce the scale of communication. If 8 GPUs are located on a single node, then we can leverage high-speed intra-node connections such as NVLink/NVSwitch to perform the most communications. Next, we give a general form of model states partitioning in our system and provide an analysis of the benefits.


In \sysname, we divide all the devices into multiple groups and partition the model states within each group. Every group has the same number of devices and holds a complete replica of the model states in training. We call these groups \emph{partition groups}. Each device is tagged with a local group rank. Devices with the same local group rank form another type of group, named \emph{replication group}, and they hold the same part of the model states. In Figure~\ref{fig:sys_design_partition}, we give an example that the model states are partitioned onto two devices. Thus every two devices with consecutive rank numbers form a partition group. The devices ranked with odd numbers and even numbers form two replication groups separately. During the training, when a parameter tensor is needed for either the forward or backward computation, \sysname invokes all-gather collective to gather the corresponding model parameters distributed within each partition group. After the gradients are computed on each device, \sysname uses all-reduce collective to aggregate the gradients, and then it partitions the gradients within each partition group.

Now we give the performance analysis of our partitioning strategy in terms of the cost of all-gather. We assume the iteration time is bounded by the communications, which is true based on our measurements (\S\ref{sec:motivation_communication_bottleneck}). The time cost of ZeRO-3's partition-to-all strategy is $C_{\text{all}} = ((n-1)M)/(n B_{\text{all}}) $, where $B_{\text{all}}$ denotes effective bandwidth among all devices. The time cost of \sysname is $C_{\text{\sysname}} = ((p-1)M)/(p B_{\text{part}}) $. Here we assume all partition groups have the same intra-group bandwidth, and we denote this bandwidth by $B_{\text{part}}$. Because the value of function $(x-1)/x$ increases when $x \ge 1$ and $p \leq n$, we have the following inequality for the ratio of two costs.
\[ \frac{C_\text{all}}{C_\text{\sysname}} \ge \frac{B_{\text{part}}}{B_{\text{all}}}.\]
For models that can be partitioned to devices located on a single node, only local high-speed NVLinks connections are used for all-gather. Based on the measurement on 64 GPUs spread across 8 computational nodes, we get $B_{\text{part}} \simeq 128\text{GB/s}$ and $B_{\text{all}} \simeq 11\text{GB/s}$. Thus, the cost ratio can be as large as 11.6. For
the case where there are 32 nodes in total and each partition group consists of 4 nodes, the cost ratio is ranging from 2.7 to 4.9 based on our measurements presented in \S\ref{sec:motivation_communication_bottleneck}. For the model states that can be partitioned within 4 nodes, we can expect about 63.6\% to 91.3\% time reduction for parameter gathering with our partitioning strategy.

\subsection{Hierarchical Communication Strategy}
\label{sec:sys_design_hierarchy_sync}

\begin{figure}[t]
  \centering
  \includegraphics[width=0.95\linewidth]{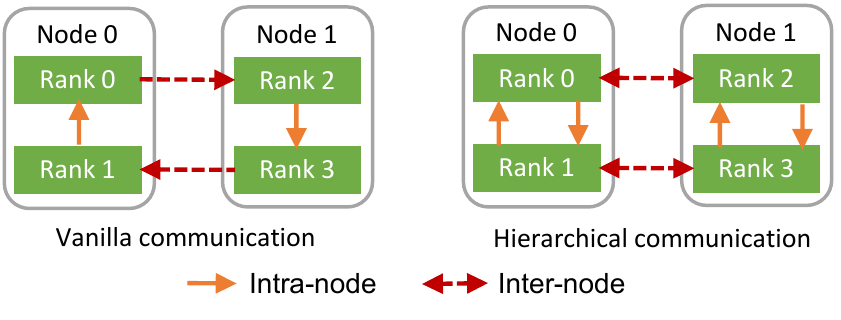}
  \caption{Differences between hierarchical communication and vanilla communication.}
  \label{fig:hier_comm_channels}
\end{figure}

When the model size grows, it requires more devices to hold the model states for the training. If the required devices span multiple computational nodes, inter-node communication is needed for parameter gathering. We can reduce the transmitted data volume over inter-node connections by reducing the scale. Assuming we want to all-gather a message with size $M$ among $p$ participants, the data traffic transmitted among participants is determined by $(p-1)M/p$~\cite{CollectiveCommunicationTheory_chan2007}. This means we can split $p$ participants into multiple small groups and perform independent communication within each group. We consider GPUs spanning across multiple nodes as a two-dimensional grid, in which we first aggregate the data across nodes in parallel and then merge the local data on each node.


For hierarchical communication to work properly, we first build communication channels for devices.
Assuming each computational node has $k$ devices, \sysname builds $k$ communication channels for inter-node communication and a separate communication channel for intra-node communication (Figure~\ref{fig:hier_comm_channels} (right)). As a comparison, vanilla collective communication uses a single communication channel for devices spanning across nodes (Figure~\ref{fig:hier_comm_channels} (left)). In Figures~\ref{fig:hier_comm_channels}, we illustrate the idea using two computational nodes, each of which has two devices, i.e., $p=4$ and $k=2$. Next, we introduce how hierarchical communication works for inter-node communication.

\begin{figure}[t]
  \centering
  \includegraphics[width=0.85\linewidth]{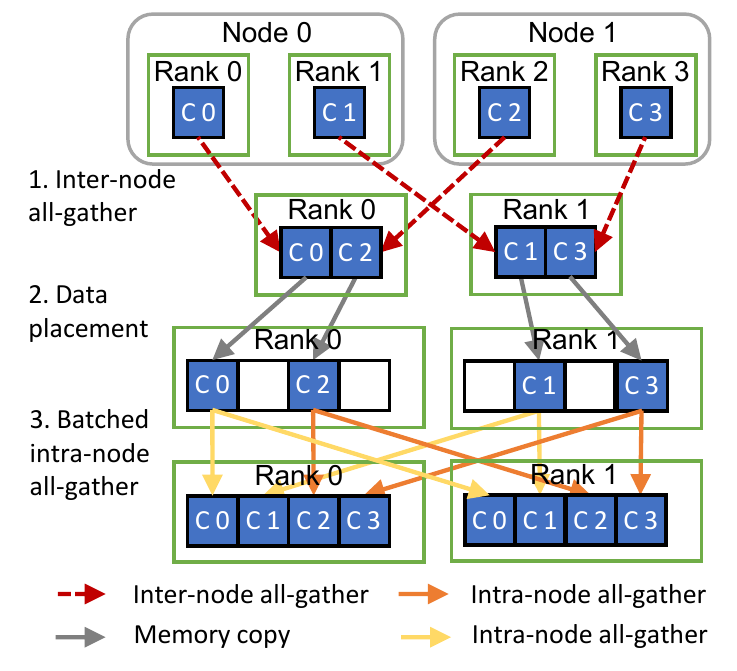}
  \caption{Hierarchical communication stages; C* denotes a data chunk; the procedure of Node 0 is shown for brevity.}
  \label{fig:hier_comm_orders}
  \label{fig:sys_design_hierarchy_comm}
\end{figure}

\sysname uses a three-stage algorithm for hierarchical communication. In the first stage, each device uses the inter-node communication channels to do all-gather with the devices that have the same local rank on respective nodes. The inter-node all-gather operations are executed in parallel. In the second stage, the data chunks are rearranged to ensure correctness. In the third stage, we invoke batched intra-node all-gather. In general, for the model states partitioned onto $p$ devices spanning $p/k$ nodes, the number of batched all-gather calls is $p/k$ in the third stage.
An example of the algorithm running across two nodes is given in Figure~\ref{fig:hier_comm_orders}. In the following, we explain why we have the second and third stages work as we described here.


The second and third stages are designed to fix the memory discontiguous issue. Otherwise, we would get the wrong output.
We use the model states partitioned to two nodes with 4 GPUs for an explanation, shown in Figure~\ref{fig:hier_comm_orders}.
The final outputs of the hierarchical communication algorithm should place
data $C0$ and $C1$ in the adjacent locations. However, the inter-node all-gather will gather $C0$ and $C2$ into a contiguous memory. Thus, if we directly launch an all-gather collective primitive on the
output from the first stage, we will get the wrong memory layout $[C0, C2, C1, C3]$, while the correct one is $[C0, C1, C2, C3]$. To fix this, we add a data movement stage before intra-node all-gather to rearrange the data chunks. Then in the third stage, we launch $p/k=2$ intra-node all-gather collectives in a batch, where each intra-node all-gather operation works on a subset of the data chunks.
Launching multiple communications in a batch requires new communication API implementation to get good performance, which is detailed in \S\ref{sec:implementation}.

The performance benefits of the hierarchical communication strategy depend on
the scale of the model states partitioning. Assume the model is partitioned onto $p$ devices, and $p$ is divisible by $k$, where $k$ is the number of devices on each computational node. With the vanilla communication strategy, the inter-node data traffic is $(p-1) M / p$. Using the proposed hierarchical communication, the data volume transmitted over inter-node connections is reduced to $(p-k) M / p$.  In this way, the communication volume over the slow inter-node links is reduced by
\[\frac{p-1}{p-k}.\]
Given that $p \ge k \ge 1$, this ratio decreases monotonically and approaches $1$ when $p$ increases. Thus, the improvement is less when we have to use more devices to hold a model replica. In a typical setup, we would have $k=8$. A $10$B-$50$B parameter model typically requires $8 \leq p \leq 64$ number of workers for holding the model states. In this case, we will obtain $11.1$\% to $46.6$\% data volume reduction with hierarchical communication.

\subsection{2-hop Gradient Synchronization}
\label{sec:sys_design_grad_sync}



In the typical distributed training setting, we need to aggregate gradients across all the devices \cite{goyal2017accurate}. Gradient aggregation is an expensive synchronization step and its cost scales with the number of workers, detailed in \S\ref{sec:motivation_communication_bottleneck}. It ensures that all devices work on the same model states.

To improve the training efficiency, more recent works advocate large-batch training~\cite{li20211,you2019large,goyal2017accurate,you2018imagenet,zheng2020accelerated}. However, due to the limited device memory, practitioners have to resort to gradient accumulation that divides a large batch into multiple micro-batches and accumulates the gradient w.r.t. each micro-batch into a shared memory buffer. In the standard data parallel setting, the gradient synchronization is only needed at the accumulation boundary where all the gradients have been computed.
However, ZeRO requires additional gradient synchronization within each micro-step because of gradient partitioning. Since each device is only responsible for holding a part of the gradient, the gradient needs to be partitioned once it is computed. In order to avoid losing the gradient information, gradients have to be aggregated before the partitioning. This makes every gradient partitioning step become a global synchronization barrier among all devices. Since \sysname only partitions the model states into a small group of devices, we can restrict the gradient synchronization within the group for each micro-step and delay the global gradient synchronization to the accumulation boundary. This motivates the design of 2-hop gradient synchronization schedule without over-paying communication costs.

\begin{figure}[t]
  \centering
  \includegraphics[width=\linewidth]{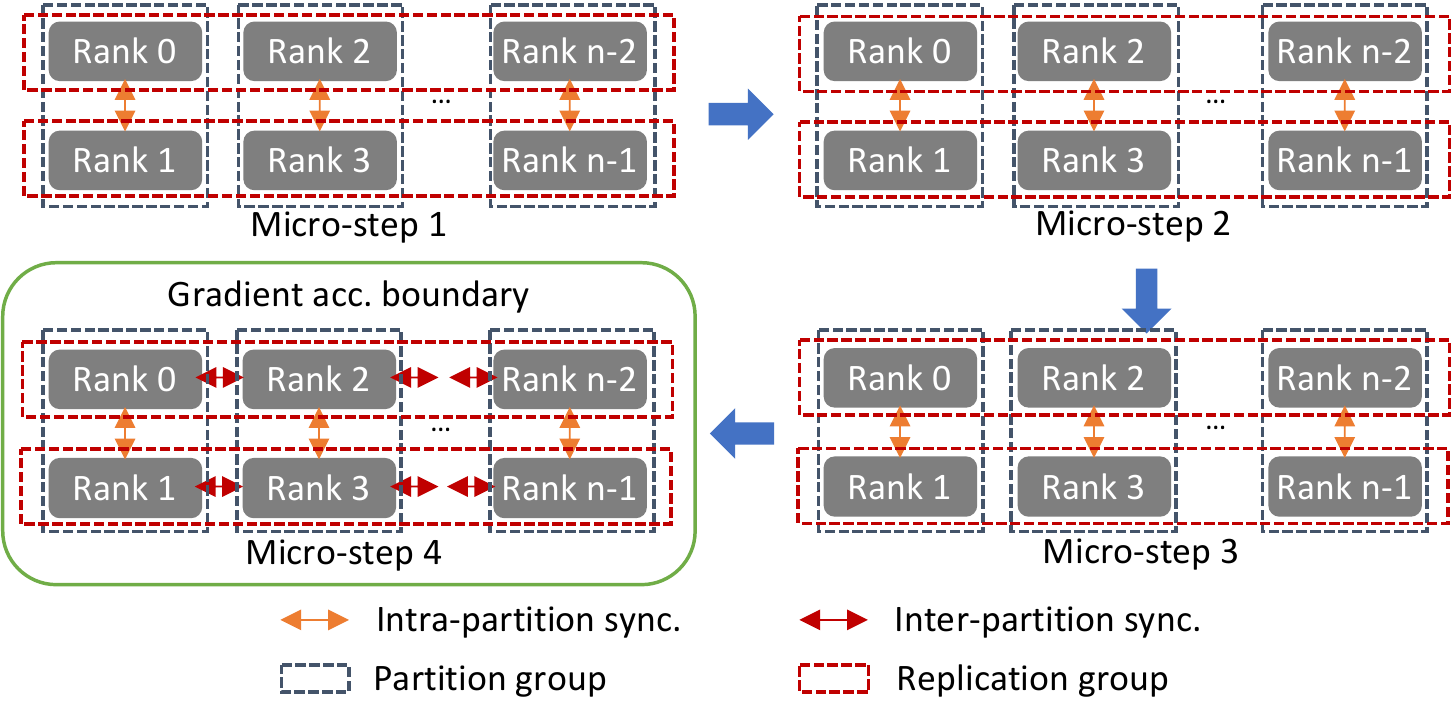}
  \caption{Gradient synchronization steps. A partition group consists of two nodes. The intra-group synchronization happens in every micro-step, while the inter-group synchronization only happens at gradient accumulation boundary.}
  \label{fig:sys_design_grad_sync}
\end{figure}

2-hop gradient synchronization performs gradient synchronization within each partition group for each micro-step. Only at the gradient accumulation boundary, global synchronization is performed among the devices that possess the same part of the model. Figure~\ref{fig:sys_design_grad_sync} gives an example of a model partitioned onto two devices. Every two consecutive ranks form a partition group. Ranks with odd number and even number indices form two different replication groups, respectively. For illustration purposes, we assume the number of gradient accumulation steps is $s=4$.
For each micro-step, \sysname uses reduce-scatter to synchronize gradients within each partition group. At the gradient accumulation boundary, an all-reduce operation is used within each replication group for synchronization. An alternative synchronization schedule is to use all-reduce for gradient synchronization at every micro-step and then partition the gradient on each device. During the partitioning, each device only keeps the part of the gradient that it is responsible for while discarding the rest. This alternative schedule is the default one implemented in DeepSpeed. However, this scheme is redundant and overpays the communication costs.

The performance benefits of the 2-hop gradient synchronization schedule depend on the number of micro-steps and the effective communication bandwidths within partition groups and replication groups. For simplicity, we assume that every partition group has the same effective bandwidth $B_{\text{part}}$, and bandwidth within each replication group is $B_{\text{repl}}$. The time cost of 2-hop schedule is $C_{\text{2-hop}} = (s M (p-1)) / (p B_{\text{part}}) + 2 M (n-p) / (n B_{\text{repl}}) $, while the time cost for the alternative schedule is $C_{\text{alt}} = 2 s M (n-1) / (n B_{\text{all}}) $. We take the ratio of two costs, and simplify the ratio using inequalities $(p-1) / p \leq (n-1)/n$ and $(n-p) / n  \leq (n-1) / n$ when $n \ge p \ge 1$. In the following inequality, we can view the right-hand side as the lower bound for the improvement.
\[
\frac{C_{\text{alt}}}{C_{\text{2-hop}}} \ge
\frac{ \frac{2 s}{B_{\text{all}}}}{ \frac{s}{B_{\text{part}}} + \frac{2}{B_{\text{repl}}}}.
\]
Assuming $s=4$, which is a reasonable setup for large batch training~\cite{megatron_lm,ms_zero}, and assuming $B_{\text{all}} = B_{\text{part}} = B_{\text{repl}}$ for simplicity, we get the lower bound of the ratio at $4/3$. This means at least 25\% cost reduction by using the 2-hop schedule. Taking heterogeneous bandwidth into consideration would further reduce the denominator on the right-hand side and helps achieve more gains. We notice that when $s=1$, under the assumption that $B_{\text{all}} = B_{\text{part}} = B_{\text{repl}}$, the 2-hop synchronization is sub-optimal compared to the alternative schedule. However, given the heterogeneity of the effective bandwidths in a large cluster, e.g., having $B_{\text{part}} \simeq B_{\text{repl}} > 1.5 B_{\text{all}}$ (which is reasonable based on our measurement in \S\ref{sec:comm_overheads}), the 2-hop schedule typically costs less. Therefore, even for $s=1$, in training large models with a large cluster, 2-hop is still preferred.

\section{Implementation}
\label{sec:implementation}

The implementation of \sysname is based on DeepSpeed-v0.4.9 and PyTorch-v1.11.
To efficiently implement our design, we make the following optimizations. 

\parabf{Fine-grained synchronization.}
Both parameter gathering and gradient synchronization involve a large number of communication kernel launches.
Communication and computation operations are typically executed asynchronously from each other in their own CUDA streams.
To maintain the data dependency correctly among these two types of operations, synchronization is required at proper position.
The synchronization mechanisms like device synchronization or stream synchronization used in DeepSpeed-v0.5.6 operate in a coarse granularity and hence lead to sub-optimal communication and computation overlapping, especially on lower bandwidth clusters. For example, if a communication operation $comm0$ depends on the output from computation operation $comp0$, and the $comp0$ is running with another computation $comp1$ on the device, then using coarse-grained device synchronization would delay $comm0$ until $comp1$ is completed.
Instead, \sysname follows the good practice in existing works, e.g., BytePS~\cite{byteps}, that leverages much finer-grained \verb+wait_event+, \verb+wait_stream+ and \verb+record_stream+ operations for synchronization, which allow us to maintain the relative order of computation and communication operations in different streams.
Using this mechanism, $comm0$ can kick off without waiting for $comp1$ to complete. In addition, during the forward and backward passes, many complex decisions need to be made, such as which parameters should be fetched, predicting which parameters will be used next, which parameters may be reused soon and should be kept, and which can be released. We observe that making these decisions on-the-fly creates large computation and communication bubbles. We optimize this computation by precomputing and caching the decisions. The same decisions are reused throughout the training. 

\parabf{Coalesced communication APIs}. \sysname's hierarchical communication design introduces multiple communications over small messages.
One way to improve bandwidth utilization is to batch communications.
However, it is suboptimal to use existing \verb+all_gather+ and \verb+reduce_scatter+ operators in PyTorch to implement batched communication as we will have to explicitly use a custom interleaving scheme to copy the tensors into a shared buffer. 
\sysname introduces two coalesced communication APIs, \verb+all_gather_coalesced+ and \verb+reduce_scatter_coalesced+.
These APIs avoid the redundant buffer allocation and memory copy in all-gather and reduce-scatter API calls in PyTorch.
\sysname leverages the \verb+group+ primitive in \verb+nccl+ to launch multiple communication operations at once, without extra data movement or allocation.

\parabf{Memory defragmentation}.
Like DeepSpeed, \sysname also requires frequent memory allocation and deallocation operations as model states are frequently gathered and scattered.
This results in serious memory fragmentation when using the dynamic allocation provided by PyTorch memory manager, causing out-of-memory errors when we try to allocate large contiguous memory buffers.
DeepSpeed allocates contiguous memory buffers for holding gradients to mitigate the fragmentation issue. But it does not consider the fragmentation problems caused by operations related to partitioned parameters and gradients.
\sysname's memory management solves the memory fragmentation issue in a more comprehensive way.
\sysname pre-allocates large contiguous memory buffers for holding partitioned parameters, partitioned gradients, and temporary small buffers ahead of the training. During training, \sysname reuses these buffers proactively, rather than relies on the memory management module in PyTorch.

\begin{table*}[t]
  \centering
  \caption{Structure of language models. BERT 10B means BERT with 10 Billion parameters, and similarly for other model names. We use a sequence length of 512 for all the models. }
  \begin{tabular}{l r r r r r}
  \toprule
  Model & Hidden size & Intermediate size & \#layers & \#Attention heads & Vocabulary size\\
  \midrule
  BERT 10B & 2560 & 10240 & 127 & 40 & 32008 \\
  BERT 15B & 2560 & 10240 & 190 & 40 & 32008 \\
  BERT 20B & 5120 & 20480 & 64  & 40 & 32008 \\
  BERT 50B & 8192 & 32768 & 62 & 40 & 32008 \\
  RoBERTa 20B & 5120 & 20480 & 62 & 40 & 50265 \\
  GPT2 20B & 5120 & 20480 & 62 & 40 & 50265 \\
  \bottomrule
  \end{tabular}
  \label{tab:eval_model_structures}
\end{table*}



\section{Evaluation}
\label{sec:evaluation}
In this section, we evaluate the following three aspects.
\begin{itemize}[leftmargin=*]
    \item \textbf{Training performance}: Does \sysname provide better throughput than the existing solutions?
    \item \textbf{Effectiveness of the design}: How does each component of the system affect the performance?
    \item \textbf{Fidelity}: Is the system carefully implemented so that the training is converging correctly?
\end{itemize}

\parabf{Setups}. We conduct all experiments on AWS. Unless specified
otherwise, we use Amazon EC2 p3dn.24xlarge instances for the evaluation. Each instance
has 8 V100 (32GB) GPUs, which are interconnected via NVLink. The
theoretical aggregated GPU interconnect bandwidth within the instance is 300 GB/s. For the
inter-node communication, p3dn.24xlarge has a 100Gbps elastic fabric adaptor
(EFA).
In addition, we have also evaluated our system on Amazon EC2 p4d.24xlarge instances, which have 8
A100 (40GB) GPUs and a 400Gbps EFA network. The software environment includes
CUDA-11.0, DeepSpeed-v0.5.6, PyTorch (customization from v1.11), Megatron-LM
(git-hash d416968), and nccl-v2.10.3.

\parabf{Metric and workloads}. We use system throughput and TFLOPS as
our main evaluation metrics. Unless
specified otherwise, we use model variants based on the BERT model~\cite{bert}. We
vary the number of transformer layers and the size of each layer to get
different model configurations. We also include two other popular language models,
RoBERTa~\cite{liu2019roberta} and GPT2~\cite{gpt2}. Table~\ref{tab:eval_model_structures} summarizes the detailed model configurations.
Other than language models, we also evaluate the performance for training WideResNet to demonstrate the generality of our system.
For the language models, the Wikipedia-en corpus is used as the training dataset. We fix
sequence length to 512 for the training. For the WideResNet model, we use synthetic data with images sized $3 \times 224 \times 224$. By default, we use a micro-batch size of 8, global-batch size of 8192, mixed-precision, and activation checkpointing in training.

\subsection{Performance}
In this section, we demonstrate the performance advantages of \sysname. First, we show the scalability of \sysname against DeepSpeed, which
is the state-of-the-art (SOTA) solution using DP with model states partitioning. The TFLOPS
numbers are also reported to show the computation utilization of each GPU.
In \S\ref{sec:scalability_400G_net}, we evaluate \sysname and DeepSpeed in a different network condition, i.e., 400Gbps network.
In \S\ref{sec:100B_model_case} we provide performance numbers of the 100B model training on a large scale. In \S\ref{sec:eval_compare_megatron}, we show \sysname can outperform
Megatron-LM-3D~\cite{megatron_lm_3d}, which is a
SOTA design for transformer-based language models that uses DP and MP.

\subsubsection{Scalability in 100Gbps Network}
\label{sec:evaluation_scalability}

\begin{figure}[t]
  \centering
  \begin{subfigure}[t]{0.45\linewidth}
    \includegraphics[width=\linewidth]{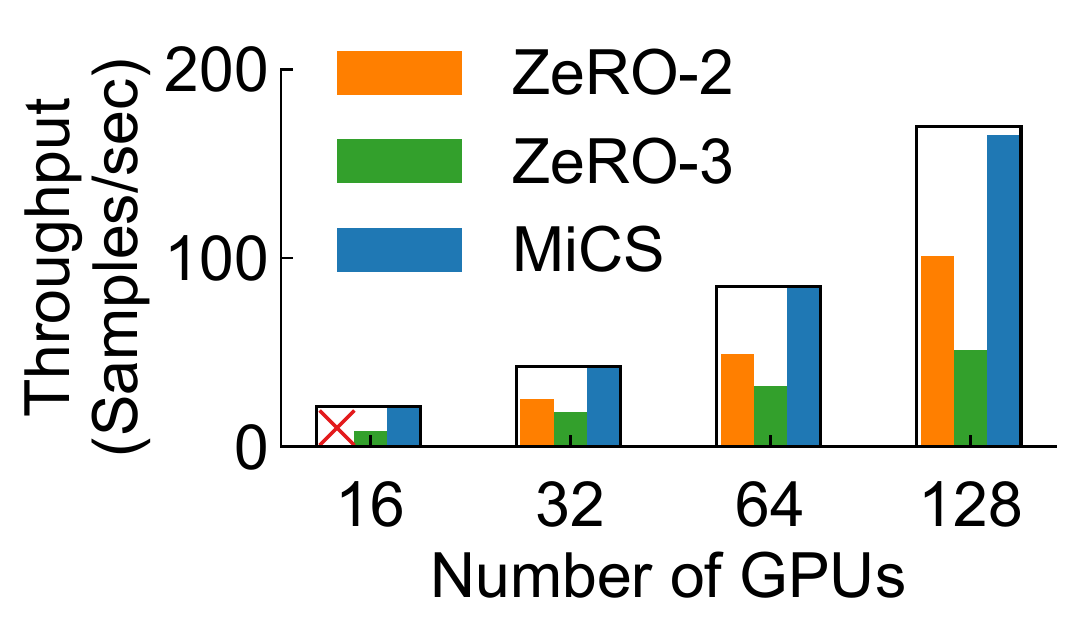}
    \caption{BERT 10B.}
  \end{subfigure}
  \begin{subfigure}[t]{0.45\linewidth}
    \includegraphics[width=\linewidth]{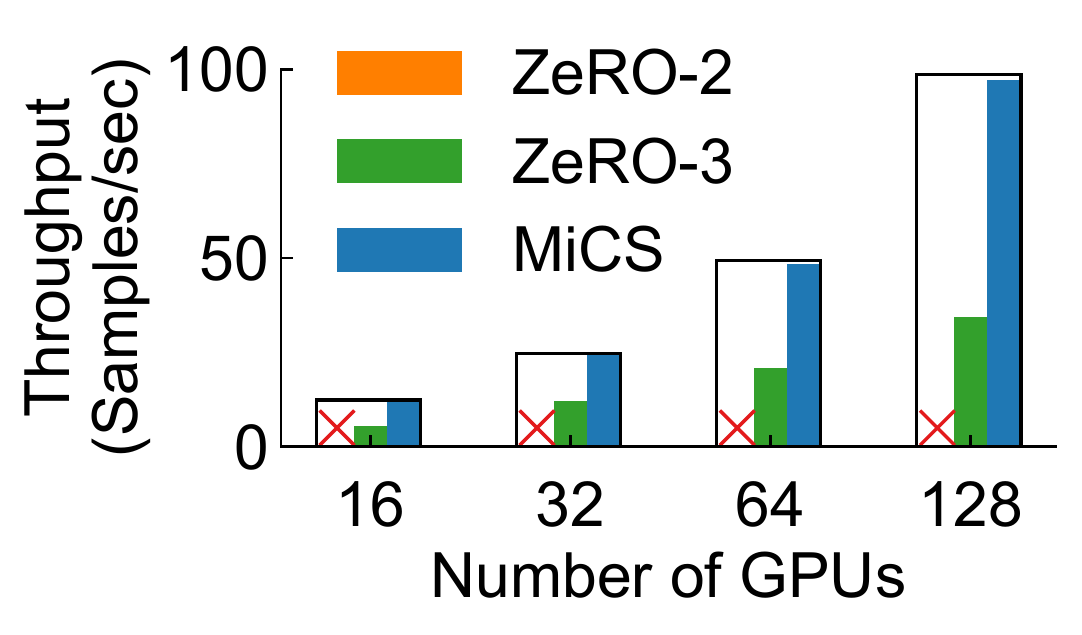}
    \caption{BERT 15B.}
    \label{fig:weak_scaling_100G_15B}
  \end{subfigure}
  \begin{subfigure}[t]{0.45\linewidth}
    \includegraphics[width=\linewidth]{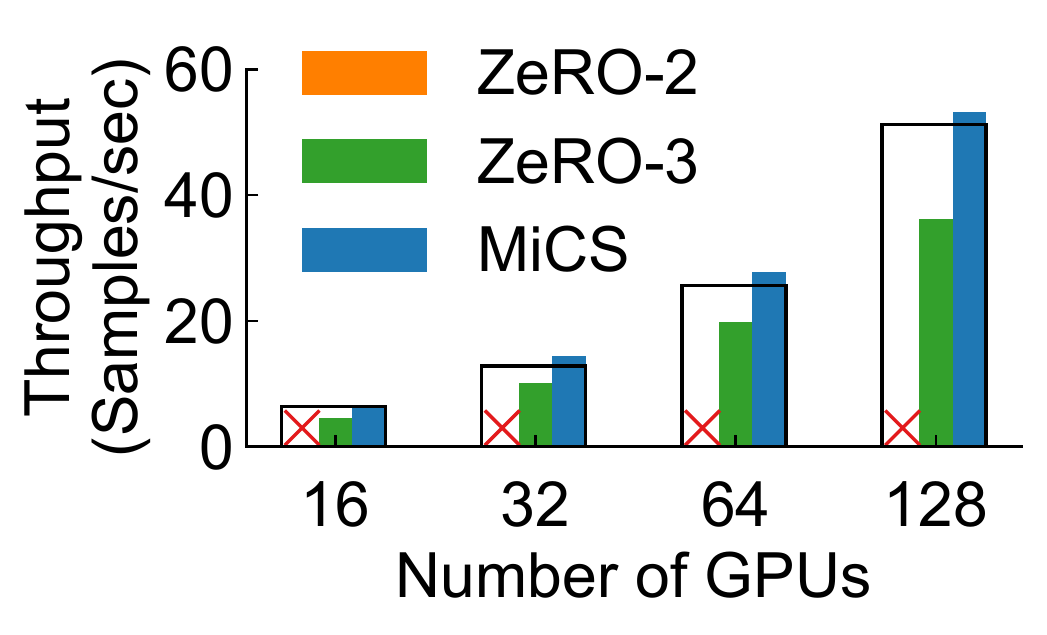}
    \caption{BERT 20B.}
  \end{subfigure}
  \begin{subfigure}[t]{0.45\linewidth}
    \includegraphics[width=\linewidth]{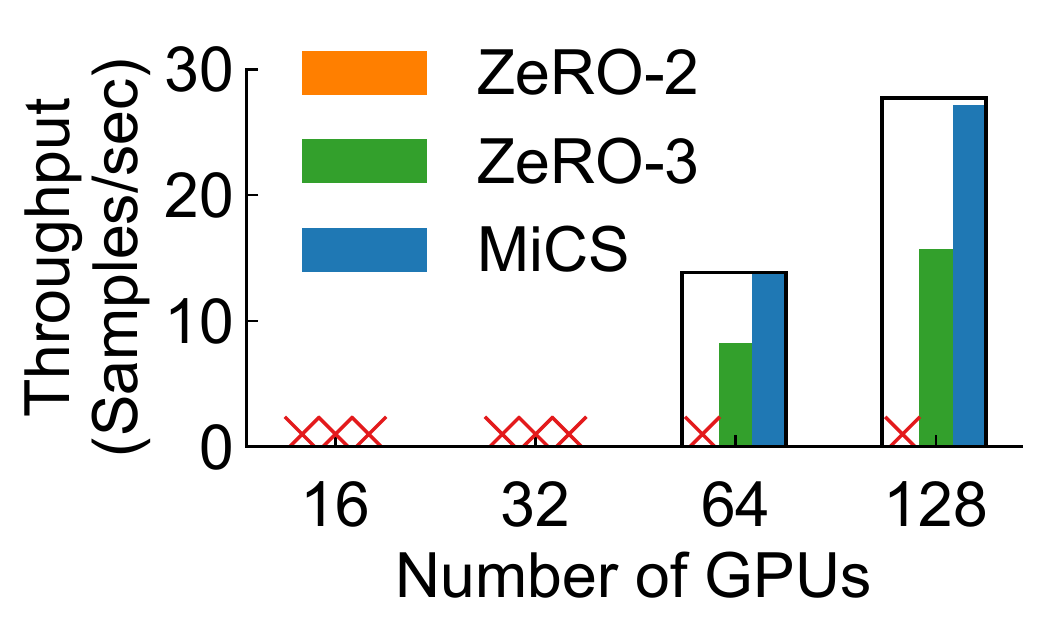}
    \caption{BERT 50B.}
  \end{subfigure}
  \caption{Strong-scaling with different model sizes; $\times$ denotes ``out-of-memory''; black rectangular denotes the linear-scaling efficiency.
  }
  \label{fig:Eval_weak_scaling_100Gbps}
\end{figure}

In this subsection, we report the throughput performance and strong-scaling efficiency of \sysname and DeepSpeed in 100Gbps networks. The baselines include both ZeRO-2
and ZeRO-3 in DeepSpeed. ZeRO-1 is excluded because it is not runnable for the smallest model we consider. ZeRO-Offload~\cite{ren2021zerooffload} and ZeRO-Infinity~\cite{rajbhandari2021zeroinfinity} are not included, either. These two variants aim to utilize CPU memory and NVMe storage to support large models, which are orthogonal to \sysname. Instead, we focus on minimizing communication overhead to improve the training throughput. For both ZeRO-3 and \sysname, we use micro-batch size 8. But for ZeRO-2 we use a smaller micro-batch size 4, because ZeRO-2 does not perform parameter partitioning and uses more GPU memory for the redundant model parameter replicas. We
vary the number of computational nodes from 2 (resp. 16 GPUs) to 16 (resp. 128 GPUs).
For the partition group size, we use the smallest number of nodes that allow us to train models with the selected batch size, i.e., 1 node for BERT 10B, 2 nodes for BERT 15B and 20B, 8 nodes for BERT 50B. All throughput numbers are averaged over 500 iterations.

\begin{figure}[t]
  \centering
  \begin{subfigure}[t]{0.45\linewidth}
    \includegraphics[width=\linewidth]{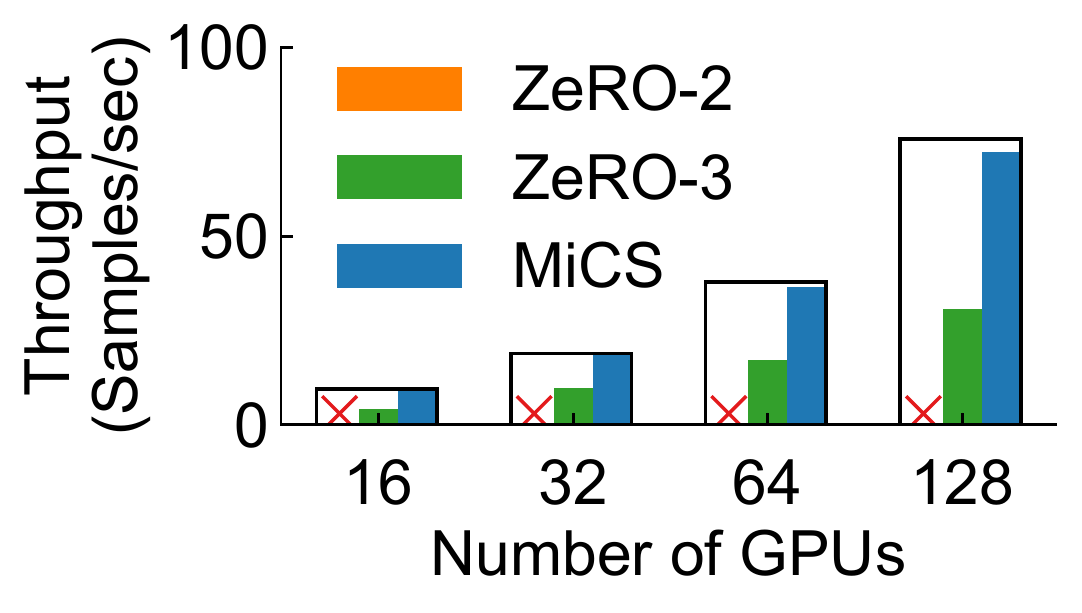}
    \caption{RoBERTa 20B.}
  \end{subfigure}
  \begin{subfigure}[t]{0.45\linewidth}
    \includegraphics[width=\linewidth]{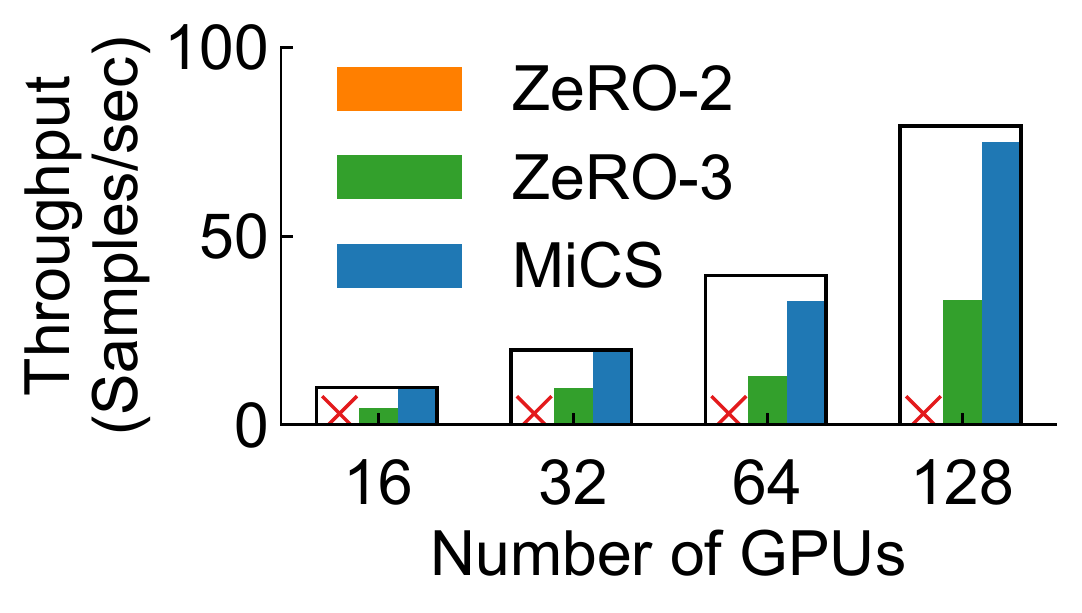}
    \caption{GPT2 20B.}
  \end{subfigure}
  \caption{Strong-scaling with other language models; $\times$ denotes ``out-of-memory''; black rectangular denotes the linear-scaling efficiency.
  }
  \label{fig:Eval_weak_scaling_100Gbps_other_LMs}
\end{figure}

As shown in Figure~\ref{fig:Eval_weak_scaling_100Gbps} and
~\ref{fig:Eval_weak_scaling_100Gbps_other_LMs}, the throughput of \sysname is significantly better than that of DeepSpeed. Our performance numbers show that the throughput of \sysname is up to 2.82$\times$ that of DeepSpeed for the BERT 15B model.
\sysname achieves near-linear or super-linear scalability in all experiments. Here we define the linear-scaling as with respect to the smallest number of computational nodes that can hold the model states with the targeted micro-batch size, e.g., for BERT 50B the linear-scaling is with respect to 8 nodes. In most of the setups, ZeRO-2 has an out-of-memory (OOM) problem. Next, we explain the rationale of our results.

The performance improvements are different with respect to the different
characteristics of the models. For the BERT 10B model, a single computational node has enough GPU memory to hold the model states so that we can leverage fast intra-node GPU interconnect to complete most of the communication. In this case, \sysname is $223$\% faster than
ZeRO-3. And, larger micro-batch size allows \sysname to further achieve more gains over ZeRO-2.
The performance gain of \sysname for BERT 15B is larger than
that for BERT 20B model. The difference is mainly due to the structural differences between
the two models. As listed in Table~\ref{tab:eval_model_structures}, BERT 15B
has narrower transformers layer but a larger number of layers. The narrower model leads to smaller computation and communication units, which allow finer grained overlapping of computation and communication.
In BERT 20B experiments, we observe super-linear scaling. This is because we have to disable hierarchical communication on 16 GPUs due to the memory constraint. 
The all-reduce overhead among replication groups is amortized by multiple micro-batches (\S\ref{sec:sys_design_grad_sync}). The amortized overhead is relatively small, less than 1\%, to the iteration time of each micro-step. Thus, MiCS can maintain near-linear scalability.

\begin{figure}[t]
  \centering
    \includegraphics[width=0.9\linewidth]{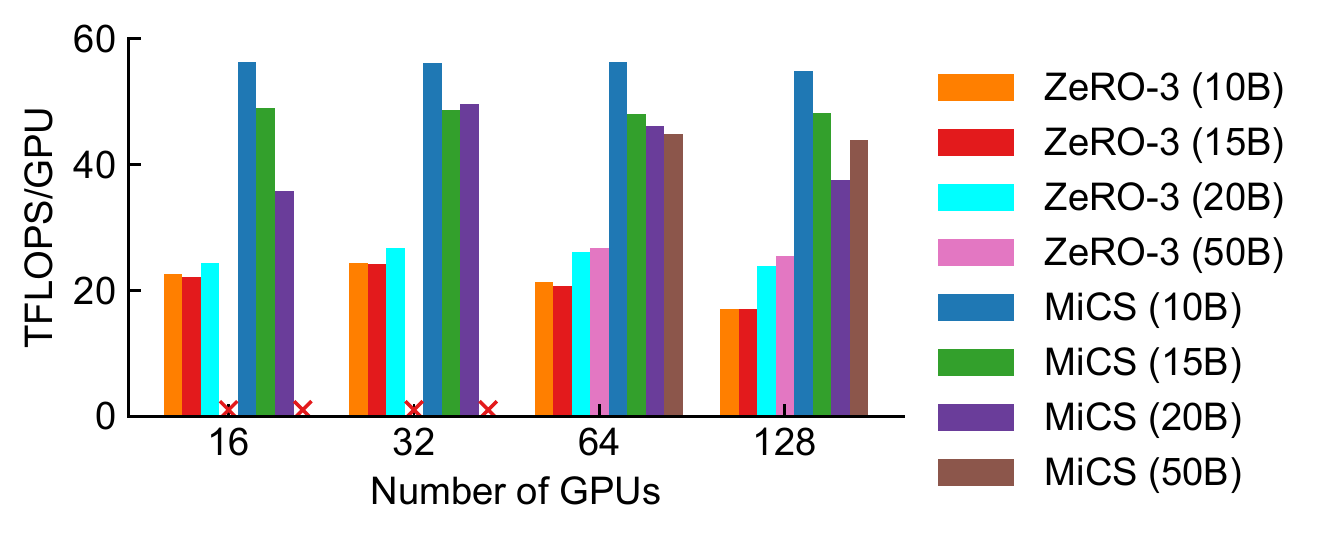}
  \caption{TFLOPS performance; BERT models with different sizes; $\times$ denotes ``out-of-memory''. }
  \label{fig:Eval_tflops}
\end{figure}

To compare the computation utilization, we calculate the TFLOPS performance based on system throughput. The TFLOPS numbers are shown in Figure~\ref{fig:Eval_tflops}. We follow the equation in \cite{megatron_lm_3d} to calculate the total TFLOPS.
\[F = 96 T l L h^2 \left(1 + \frac{l}{6h} + \frac{V}{16Lh} \right) ,\]
where $V$ denotes vocabulary size, $l$ is the sequence length, $h$ is the hidden size, $L$ refers to the number of layers, and $T$ is throughput per second\footnote{The derivation process of the formula is in the appendix of the paper \cite{megatron_lm_3d}. }.
As we can see, \sysname is better than ZeRO-3 by a large margin for all the model sizes. The maximum gain we observe is 223.7\%. For the BERT 10B model, we
achieve about 42\% of the theoretical peak performance of V100. When the model size is over 10B, the performance dropping is mainly because of the cross node
partitioning, which causes a larger communication overhead. However, the
computation utilization we get is still on par with the numbers reported by DeepSpeed
ZeRO~\cite{ms_zero} and Megatron-LM~\cite{megatron_lm} on DGX-2 clusters,
which have 800Gbps networking.


\subsubsection{Scalability in 400Gbps Network}
\label{sec:scalability_400G_net}
In this subsection, we evaluate \sysname on a GPU cluster with A100 GPUs and
400Gbps network (Amazon EC2 p4d.24xlarge instances, 8 GPUs per instance).
We use the BERT 15B and BERT 20B models for the evaluation, and we fix the micro-batch
size to 8 for all experiments. DeepSpeed ZeRO-3 is used as our baseline for comparison.

\begin{figure}[t]
  \centering
  \begin{subfigure}[t]{0.45\linewidth}
    \includegraphics[width=\linewidth]{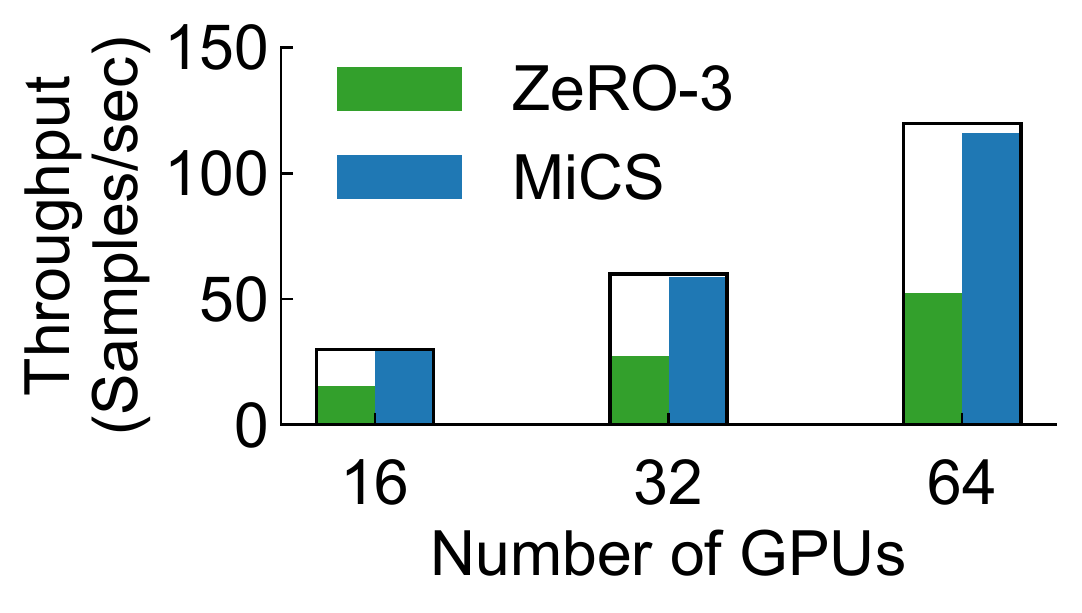}
    \caption{BERT 15B.}
  \end{subfigure}
  \begin{subfigure}[t]{0.45\linewidth}
    \includegraphics[width=\linewidth]{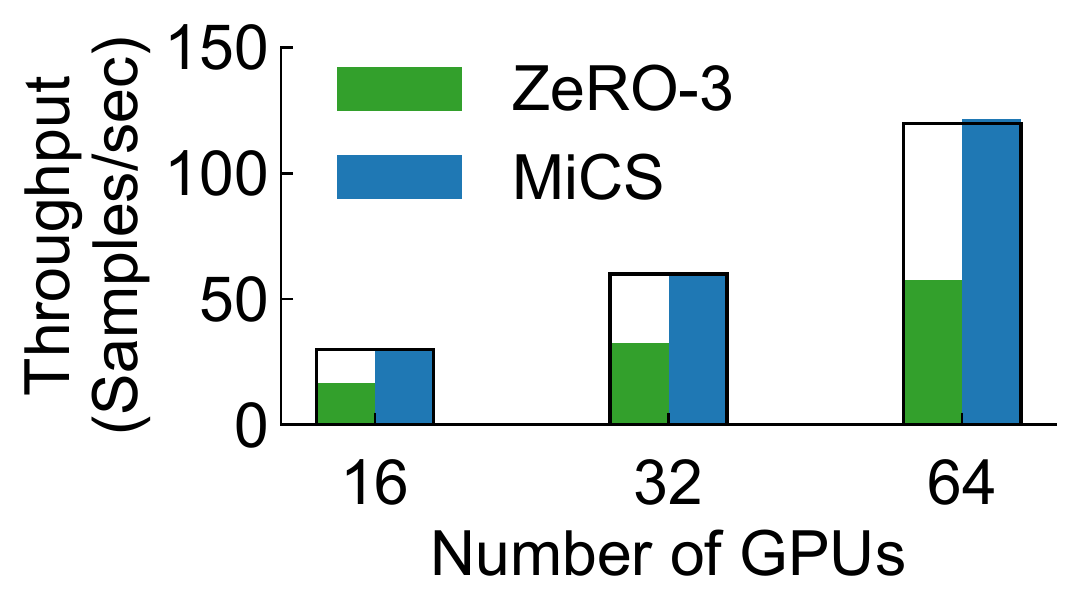}
    \caption{BERT 20B.}
  \end{subfigure}
  \caption{Throughput comparison in 400Gbps network.
  }
  \label{fig:Eval_weak_scaling_400Gbps}
\end{figure}

As shown in Figure~\ref{fig:Eval_weak_scaling_400Gbps}, \sysname significantly outperforms
DeepSpeed and achieves near-linear scaling. The throughput of
\sysname is up to 2.21$\times$ that of ZeRO-3. The throughput gap enlarges as the
scale of the cluster increases, demonstrating that
\sysname can maintain near-linear scaling efficiency. In BERT 15B case, when we scale the cluster
size from 16 GPUs to 64 GPUs, \sysname achieves 96.7\% efficiencies with respect
to 16 GPUs. In contrast, DeepSpeed ZeRO-3 only achieves 85.3\% for
BERT 15B. Compared to the results in
Figure~\ref{fig:weak_scaling_100G_15B}, the performance gains are lower mainly because faster network bandwidth mitigates communication overheads.

\subsubsection{Comparison to Megatron-LM-3D}
\label{sec:eval_compare_megatron}
We increase the number of layers to 128 while keeping the same hidden size and intermediate size as the BERT 10B model. This is because the pipeline parallelism of Megatron-LM-3D
requires the number of layers to be divisible by the size of pipeline
parallelism. We use micro-batch size 8 and global-batch size 4096 for this
experiment. We follow the takeaways from Megatron-LM-3D~\cite{megatron_lm_3d} to
tune the tensor parallel size and pipeline parallel size for better performance.
Specifically, we avoid using tensor MP across nodes and use more pipeline MP than
DP size if applicable. We report three reasonable setups of
Megatron-LM-3D, as listed in table~\ref{tab:eval_megatron_config}. In the table,
we omit the DP size, because it depends on the
size of the training cluster.

\begin{figure}[t]
  \centering
  \begin{subfigure}[t]{0.9\linewidth}
    \includegraphics[width=\linewidth]{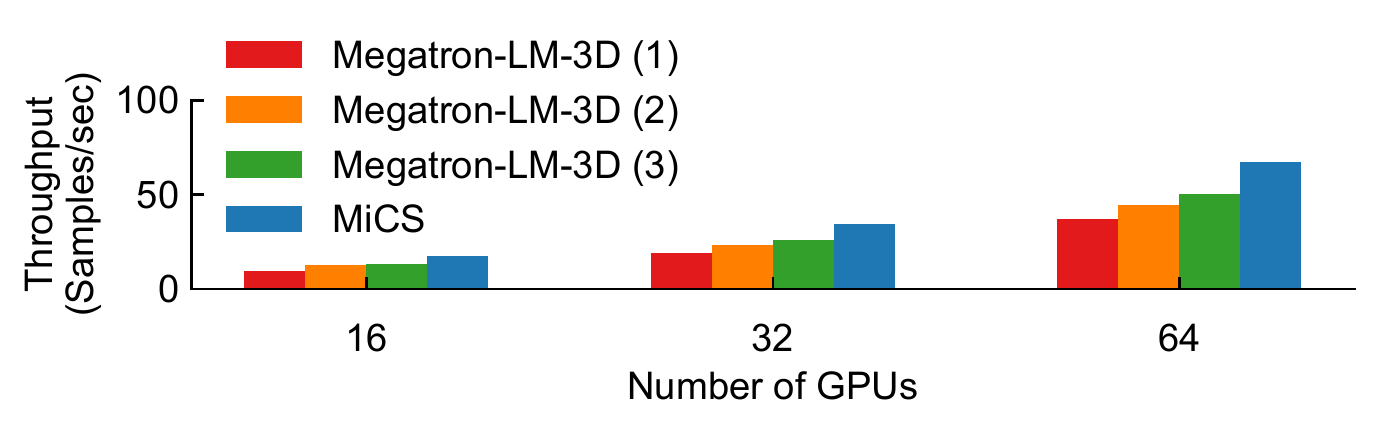}
    \caption{BERT 10B, intermediate size = 4$\times$ hidden size.}
    \label{fig:eval_compare_to_megatron_bert_10B}
  \end{subfigure}
  \begin{subfigure}[t]{0.9\linewidth}
    \includegraphics[width=\linewidth]{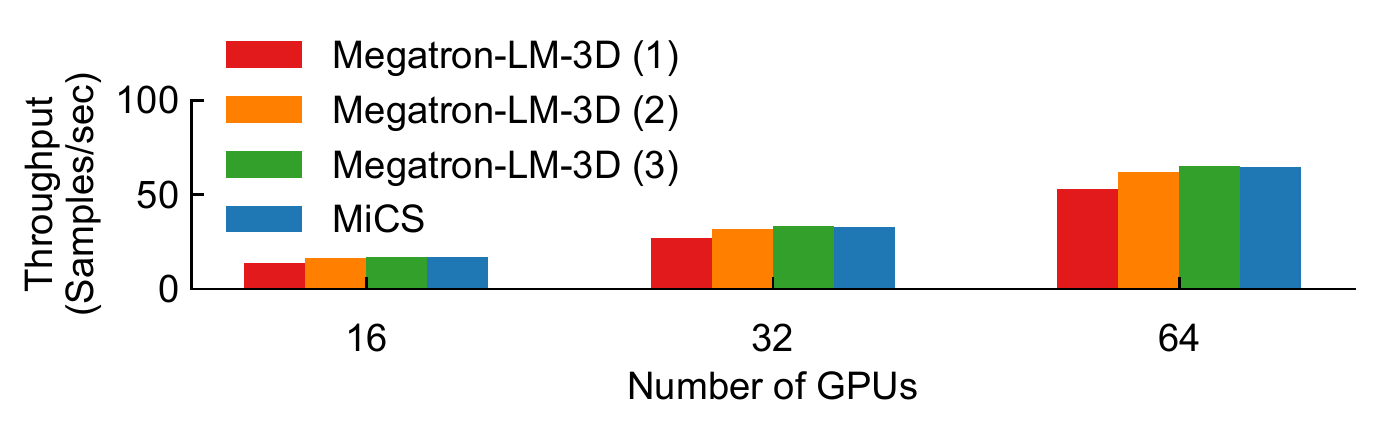}
    \caption{BERT 10B, intermediate size = 8$\times$ hidden size.}
    \label{fig:eval_compare_to_megatron_wider_8xffn}
  \end{subfigure}
  \begin{subfigure}[t]{0.9\linewidth}
    \includegraphics[width=\linewidth]{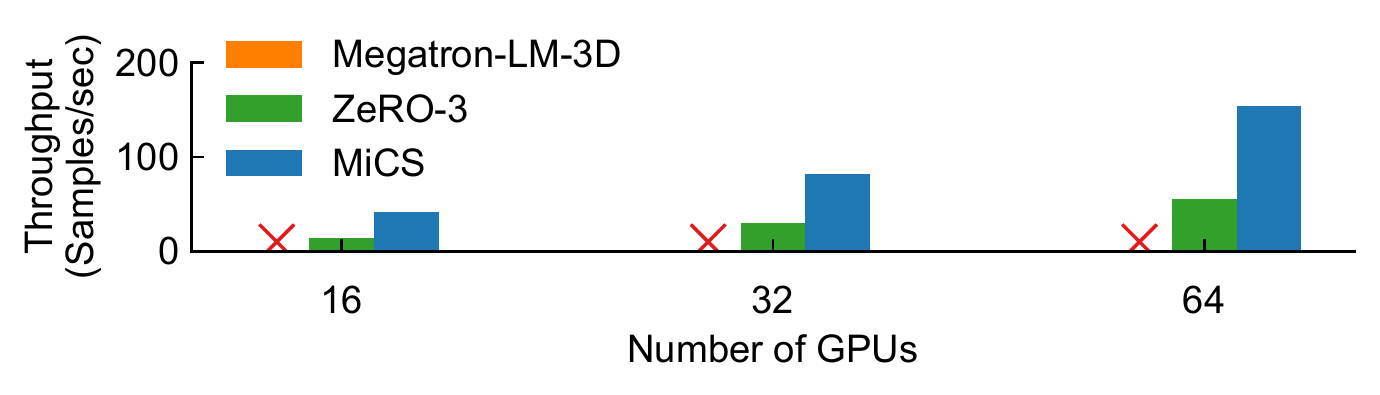}
    \caption{WideResNet 3B; $\times$ denotes ``no support''.}
    \label{fig:eval_wideresnet_3B}
  \end{subfigure}
  \caption{Performance Comparison to Megatron-LM-3D. 
  }
  \label{fig:eval_compare_to_megatron}
\end{figure}

As shown in Figure~\ref{fig:eval_compare_to_megatron_bert_10B}, the
performance of Megatron-LM-3D is sensitive to model parallel
configurations. We always restrict the tensor MP size to be lower than eight to
make sure the tensor MP ranks only communicate through NVLink. But Megatron-LM-3D is still sensitive to the tuning of the MP sizes, e.g., configuration (3) is 38\% better than configuration (1). This raises
usability challenges to users. In contrast, \sysname does not have such
complicate configurations for different parallel sizes, because of the
simplicity of data parallelism. And \sysname is up to 31\% faster than the best
results from Megatron-LM-3D. Our profiling shows the inefficiency of Megatron-LM-3D is mainly due to timeline bubbles in pipeline parallelism and communication overhead in tensor parallelism. 

\zhen{For some uncommonly structured models, Megatron-LM-3D could outperform \sysname marginally. We conducted some experiments to evaluate system performance with respect to the structural differences of models. The number of parameters of the model is fixed to 10B. Figure~\ref{fig:eval_compare_to_megatron_wider_8xffn} presents the throughput of Megatron-LM-3D and \sysname, that are evaluated on a BERT model with wider transformer layers than a regular BERT 10B model. Specifically, the model consists of 80 transformer layers. The intermediate size of each transformer layer is equal to 8$\times$ hidden size. This kind of wider structure is used in the evaluation of GSPMD~\cite{xu2021gspmd}. Usually, the intermediate size of a transformer layer is 4$\times$ that of the hidden size~\cite{transformer, megatron_lm, megatron_lm_3d,ms_zero}. The number of transformer layers, 80, is chosen to match the size of the regular BERT 10B model. The other training setups are the same as the previous experiment. In this experiment, Megatron-LM-3D with configuration (3) is slightly better than \sysname. The performance gaps are within 1.5\%. For this specific setup, the wider structure produces larger intermediate activations and requires more memory for each transformer layer. Frequently allocating and releasing large memory chunks cause allocation failure and retry at the PyTorch allocator side~\cite{torch-mem-stats}, which impacts the efficiency of overlapping computation and communication in \sysname. 
}

\begin{table}[t]
  \centering
  \caption{Megatron-LM-3D configurations. }
  \begin{tabular}{l r r r}
  \toprule
  Configuration & Tensor MP size & Pipeline MP size  \\
  \midrule
  Megatron-LM-3D (1) & 8 & 1 \\
  Megatron-LM-3D (2) & 4 & 4 \\
  Megatron-LM-3D (3) & 2 & 8 \\
  \bottomrule
  \end{tabular}
  \label{tab:eval_megatron_config}
\end{table}

\subsubsection{Performance on CV models}
\label{sec:eval:wide-resnet}
To show the performance improvements of \sysname generalize to other models,
we report the training throughput of
WideResNet~\cite{wideresnet}, a computer vision model, in
Figure~\ref{fig:eval_wideresnet_3B}. We compare \sysname against DeepSpeed
(ZeRO-3). Note that Megatron-LM-3D cannot be applied to training this model. We scale up the size of WideResNet by enlarging the width and number
of blocks of the network. In our setup, the WideResNet model has 3B parameters. It has 200 convolution layers, width factor 8, and its bottleneck block configuration is \verb+[6, 8, 46, 6]+.
We fix
batch size 8 for each GPU, and use synthetic image data with size 224x224 for benchmarking. The training uses \verb+float32+ and activation checkpointing is
disabled. The model is not runnable under ZeRO-2 optimization. The system
throughput of \sysname is up to 2.89$\times$ that of DeepSpeed (ZeRO-3).


\subsubsection{Case Study: 52B and 100B Model Training}
\label{sec:100B_model_case}

\sysname has been deployed to train proprietary models in distribution. 
Our training
cluster consists of 128 A100 GPUs with 400Gbps networking. Our results show
that we can achieve 179 and 171 TFLOPS per GPU for 52B and 100B parameter models, respectively. These are about 57\% and 55\% compute
utilization of the peak half-precision performance of A100. The utilization results
outperform the TFLOPS performance reported from
Megatron-LM-3D~\cite{megatron_lm_3d} on DGX A100 clusters with 8 InfiniBand
networking cards (1.6Tb/s)~\cite{DGX_A100}. When we increase the number of GPUs from 128 to 512, we can obtain 170 TFLOPS per GPU for the 100B parameter model with 99.4\% weak scaling efficiency, where the partition group size is 128 GPUs. When the cluster size equals the partition group size (128 GPUs), the performance improvements come from hierarchical communication and implementation optimizations. In contrast, DeepSpeed ZeRO-3 only achieves 62 TFLOPS per GPU for training a 100B parameter model on 512 GPUs with 72\% weak-scaling efficiency. In this experiment, the size of each micro-batch is 16 and the number of micro-steps is 4. The TFLOPS performance of \sysname is 2.74$\times$ that of DeepSpeed ZeRO-3 on 512 GPUs.

\subsection{Analysis of the Design}
\label{sec:eval_analysis_design}
To understand the performance contribution of each component in \sysname, we conduct
ablation tests in this section. We divide our studies into three subsections.
Each subsection corresponds to one of the three components in \S\ref{sec:sys_design}. For each experiment, we present the setups followed by
detailed results and takeaways.

\subsubsection{Analysis of Partition Group Size}
\begin{figure}[t]
  \centering
  \includegraphics[width=0.85\linewidth]{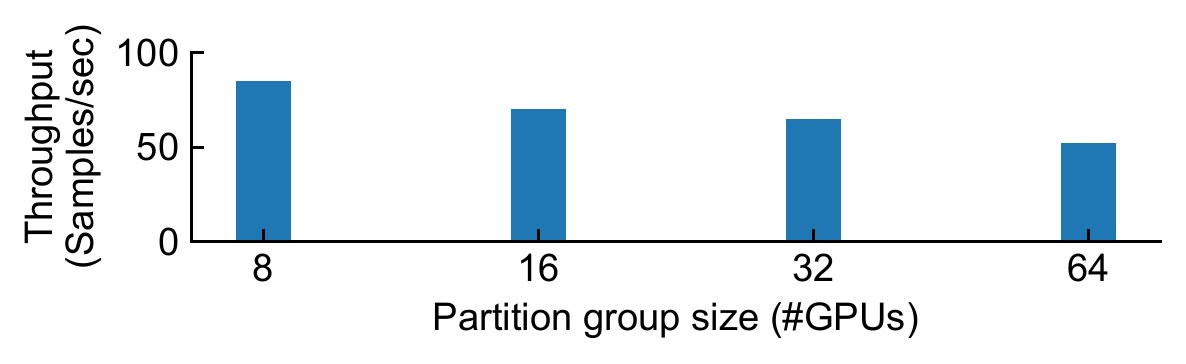}
  \caption{Throughput change w.r.t. partition group sizes.}
  \label{fig:eval_analysis_partition_size}
\end{figure}

As the scale-aware model partitioning uses partition groups for storing
model states replicas, it is natural to ask the relationship between the size of
the partition group and the end-to-end performance. In this experiment, we use BERT 10B model, fix the micro-batch size to 8, and use 64 V100 GPUs in total. We vary the size of each group from 8 GPUs to 64 GPUs. If we use all the 64
GPUs for partitioning the model states, \sysname reduces to ZeRO-3. As shown in
Figure~\ref{fig:eval_analysis_partition_size}, by increasing the partition group size, the end-to-end throughput trends down obviously. The throughput of
partition group size 8 is 1.6$\times$ that of partition group size 64. Thus,
it is preferable to partition the model states into a smallest possible group.



\subsubsection{Analysis of Hierarchical Communication}

Hierarchical communication plays an important role for good performance because it can reduce
the transmitted data volume over the inter-node connections. In this subsection,
we conduct performance analysis quantitatively to show its importance. We divide
our experiments into two parts, micro-benchmark and end-to-end training
throughput. In both experiments, we report normalized performance to baselines, i.e., vanilla all-gather and DeepSpeed ZeRO-3.

\begin{figure}[t]
    \centering

    \begin{subfigure}[t]{0.45\linewidth}
        \includegraphics[width=\linewidth]{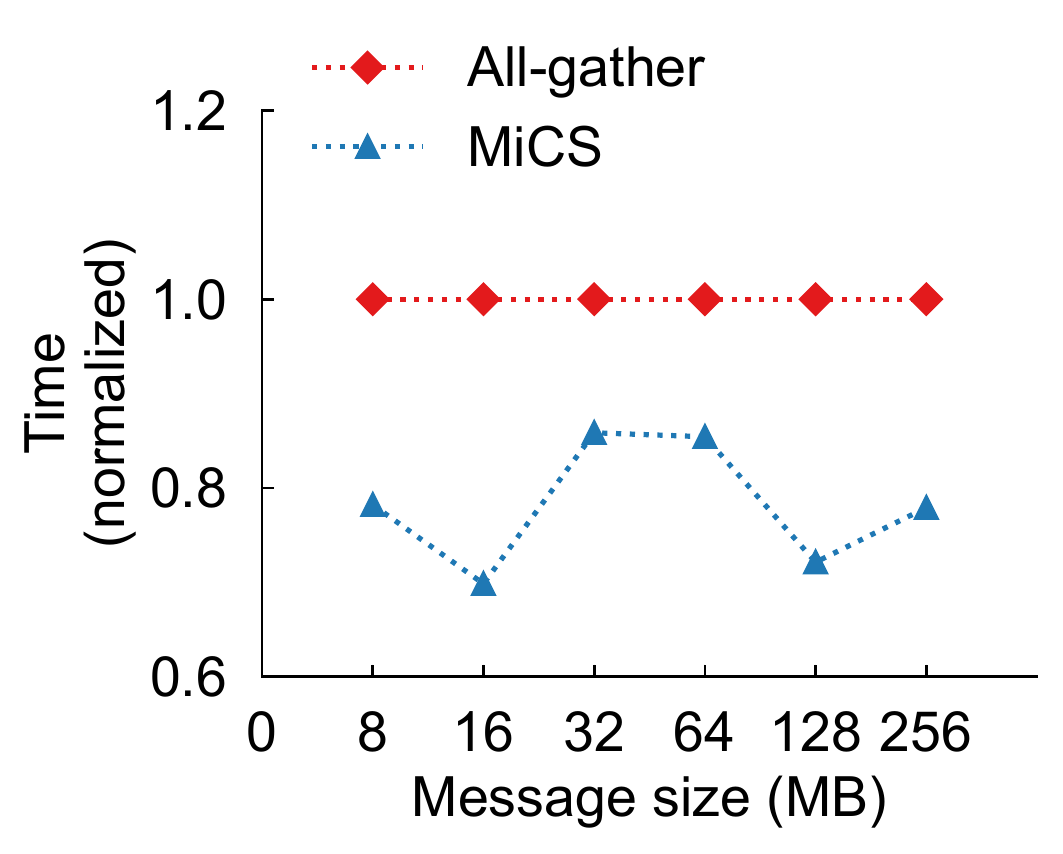}
        \caption{Micro-benchmark.}
        \label{fig:eval_hierarchical_micro_benchmark}
      \end{subfigure}
      \begin{subfigure}[t]{0.45\linewidth}
        \includegraphics[width=\linewidth]{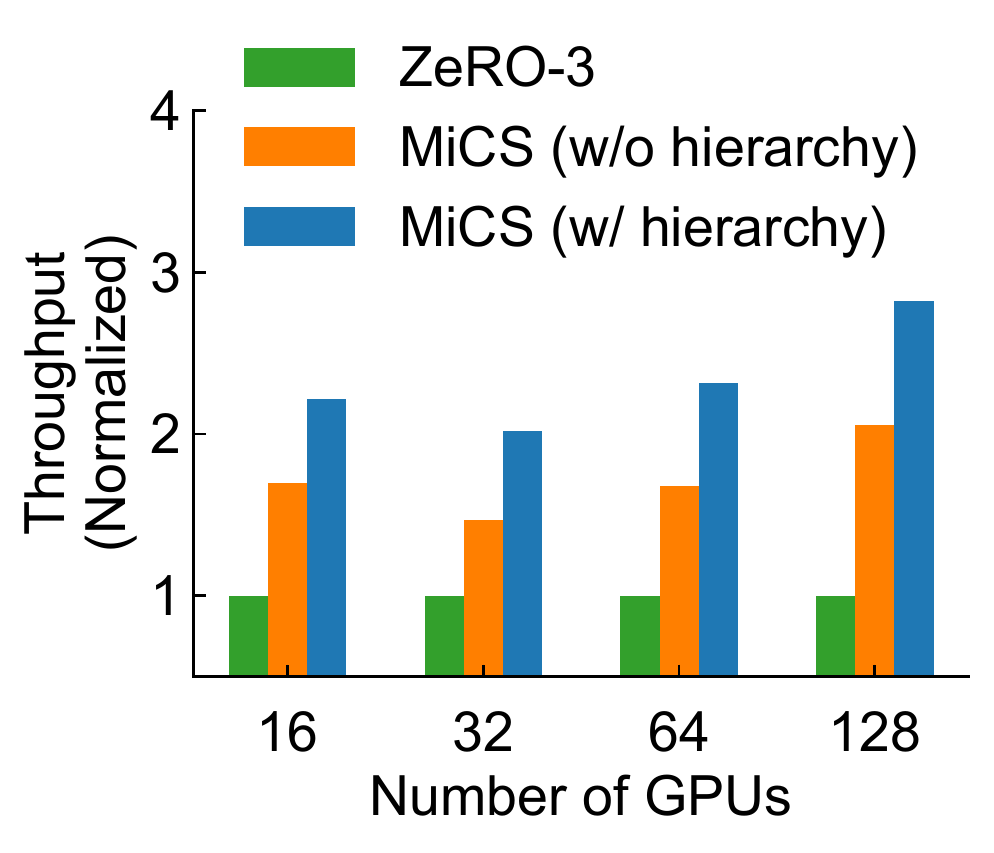}
        \caption{End-to-end performance.}
        \label{fig:eval_hierarchical_end2end}
    \end{subfigure}

    \caption{Benefits of hierarchical all-gather.
    }
    \label{fig:Eval_hierarchy_allgather}
\end{figure}

In the micro-benchmark experiment, we measure the elapsed time of vanilla all-gather and hierarchical all-gather operators for handling different message sizes. We use two Amazon EC2 p3dn.24xlarge instances.
We cap the
message size at 256MB, because a single parameter fetching typically gathers less data than it for better overlapping of computation and communication. In
Figure~\ref{fig:eval_hierarchical_micro_benchmark}, we can see that the elapsed time
of hierarchical communication operator is consistently lower than the baseline. For message size 128MB,
hierarchical communication only uses about 72.1\% of the time cost of vanilla
all-gather.

For the end-to-end experiment, we use the BERT 15B model, which needs two computational nodes (i.e.,16 GPUs) to hold the model states for the training. For models that can be held by a single computational node (i.e., 8 GPUs), the hierarchical all-gather is not needed. We vary
the cluster size from 16 to 128 GPUs and evaluate \sysname with and without
hierarchical communication. We normalize throughput numbers to
the results of DeepSpeed ZeRO-3. As shown in Figure~\ref{fig:eval_hierarchical_end2end}, \sysname with hierarchical communication is consistently better than the case where hierarchical communication is disabled. In particular, hierarchical communication improves the end-to-end training throughput by $30.6$\% to $38$\%.


\subsubsection{Analysis of Synchronization Scheduling}
\begin{figure}[t]
  \centering
  \includegraphics[width=0.85\linewidth]{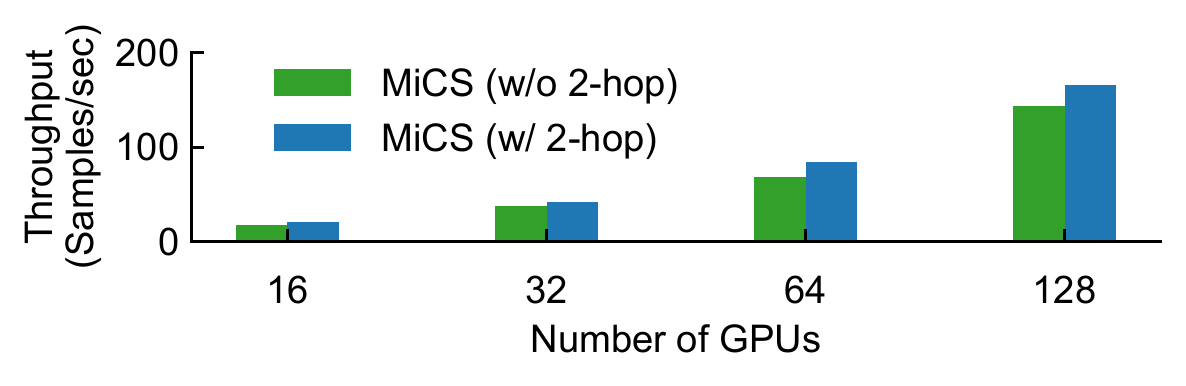}
  \caption{Benefits of 2-hop gradient synchronization. 
  }
  \label{fig:eval_analysis_sync_schedule}
\end{figure}

In this experiment, we report the throughputs of \sysname with 2-hop gradient synchronization enabled and disabled.
We use the BERT 10B model for the experiments and
fix the micro-batch size 8, global batch size 8192 for training. We partition model states
on 8 GPUs. When the 2-hop synchronization is disabled, the system uses an alternative synchronization schedule that synchronizes the
gradients across all devices at the end of each micro-step, explained in \S~\ref{sec:sys_design_grad_sync}.
We can see the performance gaps between these two setups, Figure~\ref{fig:eval_analysis_sync_schedule}. When the cluster size increases to 128 GPUs, we get the max throughput gap. Numerical
results indicate that the relative improvement ranges from 11\% to 24.9\%, when 2-hop synchronization is enabled.


\subsection{Other Optimizations}
We conduct experiments to analyze the performance improvements by using the optimization techniques described in \S\ref{sec:implementation}. We use the BERT 10B model for the evaluation. In the training, we use the default setup as mentioned at the beginning of
\S\ref{sec:evaluation}. When we turn off optimizations that are unique to \sysname and
let the model states be partitioned over all devices, \sysname reduces to ZeRO-3 with the optimization techniques in \S\ref{sec:implementation}, We denote it as ``\sysname (ZeRO-3)''. For comparisons, we report
the throughput of DeepSpeed ZeRO-3.

\begin{figure}[t]
  \centering
  \includegraphics[width=0.85\linewidth]{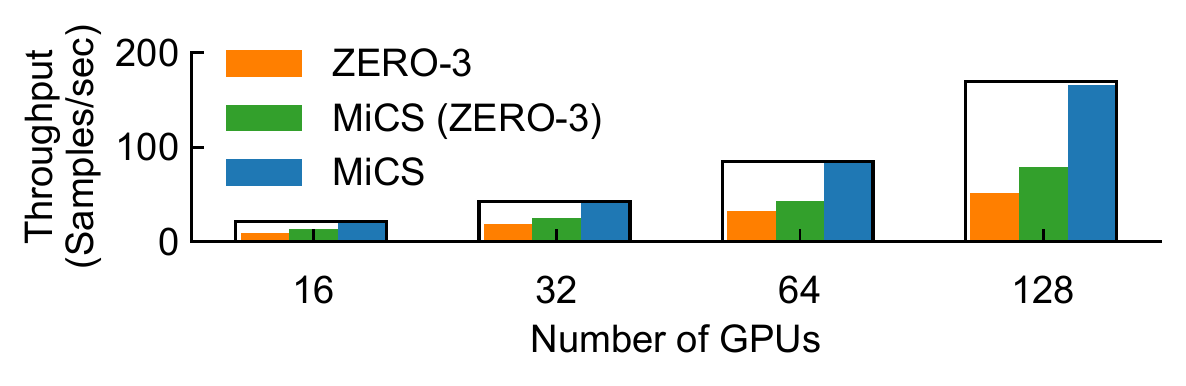}
  \caption{Improvements of implementation optimizations. 
  }
  \label{fig:Eval_ablation_implementation}
\end{figure}

Figure~\ref{fig:Eval_ablation_implementation} shows the improvements of
using the proposed system optimizations. \sysname (ZeRO-3) achieves 54.1\% better system
throughput than DeepSpeed ZeRO-3 when the cluster scales up to 128 GPUs, while the
scaling efficiency of DeepSpeed ZeRO-3 drops when we scale out the cluster. In addition, \sysname still significantly outperforms \sysname (ZeRO-3), demonstrating the superiority of minimizing the communication scale.

\subsection{Fidelity}
\label{sec:fidelity_check}
\begin{figure}[t]
    \centering
    \includegraphics[width=0.85\linewidth]{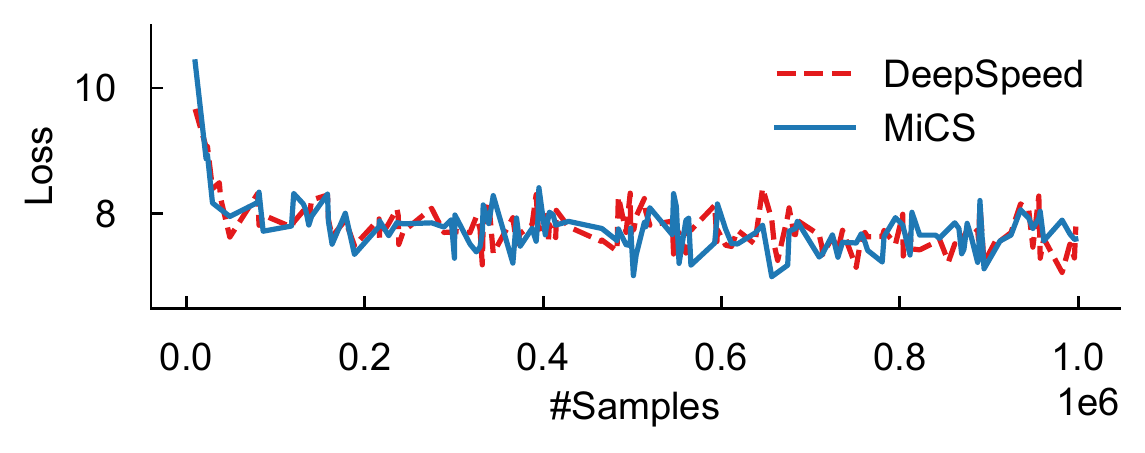}
    \caption{Fidelity of the implementation.}
    \label{fig:Eval_fidelity}
\end{figure}

In this section, we show that \sysname achieves consistent convergence as DeepSpeed, which validates the correctness of our system.
We provide the training loss curves for training a 1.5B parameter model on the Wikipedia-en dataset. The model has
48 transformer layers, each of which is constructed with the hidden size
1,600 and intermediate size 6,400. The global batch size is 512. And the micro-batch size is 8 (the number of gradient accumulation steps is 4). The loss validation process does not aim to produce exactly the same loss as DeepSpeed but to ensure the convergence behaviours are the same. We report the training losses on 1 million sequences. As shown in Figure~\ref{fig:Eval_fidelity}, \sysname provides the same convergence as DeepSpeed.


\section{Related Work}
\label{sec:related_work}

\parabf{Data parallelism.} PyTorch-DDP~\cite{pytorch},
Horovod~\cite{sergeev2018horovod}, ps-lite~\cite{ps-lite}, Tensorflow-DDP~\cite{tensorflow}, and BytePS~\cite{byteps} are
distributed training frameworks using data parallelism. All of them
place complete model states on each GPU for training. Thus, the supported model size is limited. 
Recently, ZeRO~\cite{ms_zero} has been proposed
to address the memory limitation issue of the traditional data-parallel
strategy, by partitioning the model states onto all GPUs. ZeRO-Offload~\cite{ren2021zerooffload} and ZeRO-Infinity~\cite{rajbhandari2021zeroinfinity}
are two extensions to ZeRO that explore the possibility to extend the memory to hold the model from GPU memory to CPU memory and NVMe, which are orthogonal to \sysname.
\sysname focuses on minimizing the communication overheads of the training system, addressing the challenges not solved in ZeRO.

\parabf{Other parallelisms.} ColocRL~\cite{mirhoseini2017device} 
formulates the distributed training as a placement optimization problem to
maximize the throughput. FlexFlow~\cite{flexflow} and OptCNN~\cite{jia2018exploring} use heuristic search for parallel strategies including tensor MP and device placement MP. 
Alpa~\cite{zheng2022alpa} and Unity~\cite{unger2022unity} use hierarchical search to jointly optimize within- and between-device to look for a good parallel strategy.
These systems do not explicitly embed the memory constraints into
their optimization objective, and are not directly verified to train models at the scale that MiCS trained. 
Megatron-LM-3D~\cite{megatron_lm_3d}, GPipe~\cite{huang2019gpipe}, and DAPPLE~\cite{fan2020dapple} use pipeline parallelism to
partition large models into multiple stages for the training in a synchronous manner. These solutions have resource under-utilization problems due to pipeline bubbles. PipeMare~\cite{pipemare_mlsys21}, PipeDream~\cite{narayanan2019pipedream}, and PipeDream-2BW~\cite{pipedream-2bw} use asynchronous and bounded-staleness training for efficient resource utilization which, however, \zhen{can affect the convergence quality~\cite{recht2011hogwild, de2015taming, analysis-DAWNBench}. The research direction of asynchronous methods are orthogonal to \sysname.} Currently, \sysname uses synchronous training and it does not suffer from convergence issues. DLRM~\cite{DLRM19} and Megatron-LM~\cite{megatron_lm} are specific designs for recommendation models and transformers, respectively. DLRM partitions the embedding table along row and column dimensions. Megatron-LM introduces tensor parallelism to parallelize the tensor computation on multiple devices. Megatron-LM-3D~\cite{megatron_lm_3d}
integrates the pipeline parallelism into Megatron-LM for further scaling up the
model size. Pipeline parallelism, 
tensor parallelism, and the mixture of multiple parallelisms require significantly
additional efforts to program the customized model implementation and tune the
hyper-parameters for high performance. \sysname is orthogonal to this line of research.
We compared \sysname against Megatron-LM-3D~\cite{megatron_lm_3d} in
Section~\ref{sec:eval_compare_megatron}. 

\label{sec:related-comm-optim}
\parabf{Communication optimizations.} ByteScheduler~\cite{peng2019generic} and P3~\cite{jayarajan2019priority}
overlap the computation with communication to hide the communication
cost. SwitchML~\cite{switchML} and ATP~\cite{atp} use the programmable switches as gradient
aggregation servers to reduce the communication overheads. Lossy compression
algorithms like 1bit-SGD~\cite{1bit_sgd} and DGC~\cite{lin2017deep} compress the data transmitted over
the network to improve the system performance. Those techniques are complementary to our system for further
reducing the communication overheads. Blink~\cite{wang2020blink} leverages multiple
communication channels with optimized spanning trees to speed up the gradient
synchronization. Plink~\cite{luo2020plink} discovers and explores the locality of the
distributed training cluster for better performance. Cloud Collective~\cite{luo2021cloud}
reorders the ranks of cluster nodes to explore a better topology. Blueconnect~\cite{cho2019blueconnect} decomposes all-reduce primitive with pipelined
reduce-scatter and all-gather. These techniques explore better
locality or pipeline multiple communication primitives to speed up the
synchronization. \sysname reduces communication overheads from a different perspective. In particular, our system reduces communication costs by reducing the scale of communications. Varuna~\cite{VarunaScalableLowcost2021} works on optimizing network jitter and instability among cheap ``spot'' instances~\cite{azure-spot-instance} to lower the training cost. The objective of Varuna is orthogonal to \sysname. 
\zhen{In high-performance computing, innovations~\cite{biberman2012optical, barker2005feasibility, singla2014high} at the hardware level are critical to the efficiency of communications. On the other hand, researchers explore the relationship among collective communication algorithms, software implementations, and message sizes to optimize each individual communication primitives~\cite{almasi2005optimization, thakur2005optimization}. These efforts are orthogonal to \sysname. Our system is built with GPU-aware library NCCL~\cite{nccl}.}






\section{Discussion and Future Work}
\label{sec:limit}
The optimality of the training throughput depends on the model structure, input data, and hardware. For the models used in the evaluation, we do not prove that \sysname is the optimal solution. 
\zhen{
MiCS is a pure data-parallel training system, which admittedly covers a limited space of parallelism strategies. Thus, for some less common model structures, e.g., wider feedforward layer in transformer blocks, Megatron-LM-3D could outperform \sysname in certain configurations marginally, shown in \S\ref{sec:eval_compare_megatron}. It is worth noting that adapting tensor model parallelism and pipeline parallelism requires refactoring model implementations~\cite{ms_zero}, thus is less favorable to practitioners. \sysname, as a pure DP solution, achieves state-of-the-art performance in training standard transformer-based models with billions of parameters and trades off performance for lower complexity in some less common cases. 
}

\zhen{Despite the model states replications created in \sysname, our system does not require additional hardware resources as compared to the existing ZeRO system. Partitioning one model states replication across all devices, as the existing ZeRO system does, underutilizes the memory of each device. As discussed in the first paragraph of \S\ref{sec:sys_design_topo_aware}, the memory capacity of eight V100 (32GB) GPUs are large enough for holding model states of a model with 10 billion (B) parameters. For a cluster with 16 or more V100 (32GB) GPUs, partitioning the 10B model to all devices consumes less than 32\% memory usage of each device for holding the model states. \sysname effectively leverages these spare memory resources for lowering communication costs (\S\ref{sec:sys_design}). For models that require all devices to hold the model states for training, \sysname still outperforms the ZeRO system because of the hierarchical communication module (\S\ref{sec:sys_design_hierarchy_sync}).}

\zhen{
    In \sysname, the memory consumption of each device is controlled by the size of the partition group, which is configurable. \sysname uses a heuristic to pick the size for holding the model states, which is mentioned in \S\ref{sec:evaluation_scalability}. Compared to prior large model training systems, \sysname does not introduce extra issues in terms of system practicability. For ZeRO systems, users have to figure out the smallest size of the cluster for training, otherwise the system runs into out-of-memory issues. Similarly, the Megatron-LM-3D system requires users to configure the number of pipeline stages and the tensor parallelism size, so that the partitioned model components can fit into each GPU in a cluster. In \sysname, the way to figure out the partition group size is the same as figuring out the smallest size of the cluster for training in ZeRO systems.
}

\zhen{
    To automate the configuration search for large model training, an accurate estimation of memory usage is needed. A profiling-based method can get relatively precise memory usage statistics. But once the training processing runs into the out-of-memory issue during configuration search, the dangling process can cause hanging and prevent successive configurations from launching. Estimating memory consumption from the model structure and input size is inaccurate due to the dynamic behavior of the memory management module in PyTorch runtime. Addressing challenges from estimating or predicting the memory usage of large models is beyond the scope of this paper, in which we focus on reducing the communication overheads in the ZeRO DP algorithms. 
    We leave the configuration search for \sysname as future work.
}

\section{Conclusion}
\label{sec:conclusion}

In this paper, we present \sysname, a system that attains high training throughput and near-linear scalability on the cloud by only using data parallelism. 
The overarching goal of \sysname is to minimize the communication scale so as to reduce the expensive communication overhead rooted in parameter gathering and gradient synchronization. 
Specifically, we propose scale-aware model partitioning, hierarchical communication strategy, and 2-hop gradient synchronization to achieve this goal.
We evaluate \sysname on various training workloads on large-scale clusters. \sysname outperforms DeepSpeed ZeRO by up to $2.89\times$ and demonstrates \zhen{near-linear scaling efficiency in various training setups.}



\begin{acks}
We sincerely thank the
anonymous reviewers for their valuable feedback.
We thank the Amazon Search M5 team for providing large clusters.
Xin Jin and Shuai Zheng are the corresponding authors.
Xin Jin is with the Key Laboratory of High Confidence Software Technologies (Peking University), Ministry of Education. Zhen Zhang is supported in part by NSF grants CNS-1813487 and CCF-1918757. Xin Jin is supported in part by National Natural Science Foundation of China under the grant number 62172008 and National Natural Science Fund for the Excellent Young Scientists Fund Program (Overseas).
\end{acks}


\balance
\bibliographystyle{ACM-Reference-Format}
\bibliography{xin}
\clearpage

\end{document}